\documentclass[aip,prl,reprint]{revtex4-2} 
\usepackage[T1]{fontenc}
\usepackage{times}
\usepackage{graphicx}
\usepackage{booktabs}

\usepackage{amsfonts, amsmath,amssymb, bm, graphicx}
\usepackage{siunitx}
\DeclareSIUnit{\belmilliwatt}{Bm}
\DeclareSIUnit{\dBm}{\deci\belmilliwatt}

\usepackage{amssymb}
\usepackage[english]{babel}

\newcommand{\figref}[2]{\hyperref[#1]{\ref{#1}(#2)}}
\newcommand{\figrefsub}[3]{\hyperref[#1]{\ref{#1}(#2)#3}}

\usepackage{physics}
\usepackage{lipsum}
\usepackage{changes}

\usepackage[colorlinks=true,allcolors=blue]{hyperref}
\usepackage{footmisc}

\usepackage{upgreek}

\usepackage{soul}

\makeatletter
\let\ORIbbl@fixname\bbl@fixname
\def\bbl@fixname#1{%
  \@ifundefined{languagealias@\expandafter\string#1}
    {\ORIbbl@fixname#1}
    {\edef\languagename{\@nameuse{languagealias@#1}}}%
}
\newcommand{\definelanguagealias}[2]{%
  \@namedef{languagealias@#1}{#2}%
}
\makeatother

\definelanguagealias{en}{english}

\begin{document}

\title{Nonreciprocal spin-wave dispersion in magnetic bilayers}

\author{C. Heins}
\affiliation{Helmholtz-Zentrum Dresden--Rossendorf, Institut f\"ur Ionenstrahlphysik und Materialforschung, D-01328 Dresden, Germany}
\affiliation{Fakult\"at Physik, Technische Universit\"at Dresden, D-01062 Dresden, Germany}

\author{V. Iurchuk}
\affiliation{Helmholtz-Zentrum Dresden--Rossendorf, Institut f\"ur Ionenstrahlphysik und Materialforschung, D-01328 Dresden, Germany}

\author{O. Gladii}
\affiliation{Helmholtz-Zentrum Dresden--Rossendorf, Institut f\"ur Ionenstrahlphysik und Materialforschung, D-01328 Dresden, Germany}

\author{L. K\"orber}
\affiliation{Helmholtz-Zentrum Dresden--Rossendorf, Institut f\"ur Ionenstrahlphysik und Materialforschung, D-01328 Dresden, Germany}
\affiliation{Fakult\"at Physik, Technische Universit\"at Dresden, D-01062 Dresden, Germany}
\affiliation{Radboud University, Institute of Molecules and Materials, Heyendaalseweg 135, 6525 AJ Nijmegen, The Netherlands}

\author{A. K\'akay}
\affiliation{Helmholtz-Zentrum Dresden--Rossendorf, Institut f\"ur Ionenstrahlphysik und Materialforschung, D-01328 Dresden, Germany}

\author{J. Fassbender}
\affiliation{Helmholtz-Zentrum Dresden--Rossendorf, Institut f\"ur Ionenstrahlphysik und Materialforschung, D-01328 Dresden, Germany}
\affiliation{Fakult\"at Physik, Technische Universit\"at Dresden, D-01062 Dresden, Germany}

\author{K. Schultheiss}
\affiliation{Helmholtz-Zentrum Dresden--Rossendorf, Institut f\"ur Ionenstrahlphysik und Materialforschung, D-01328 Dresden, Germany}

\author{H. Schultheiss}\email{h.schutheiss@hzdr.de}
\affiliation{Helmholtz-Zentrum Dresden--Rossendorf, Institut f\"ur Ionenstrahlphysik und Materialforschung, D-01328 Dresden, Germany}
%\affiliation{Fakult\"at Physik, Technische Universit\"at Dresden, D-01062 Dresden, Germany}

\date{\today}

%%%%%%%%%%%%%%%%%%%%%%%%%%%%%%%%%%%%%%
%%%%%%%%% Abstract
%%%%%%%%%%%%%%%%%%%%%%%%%%%%%%%%%%%%%%

\begin{abstract}
Nonreciprocal spin-wave propagation in bilayer ferromagnetic systems has attracted significant attention due to its potential to precisely quantify material parameters as well as for applications in magnonic logic and information processing. In this study we investigate the nonreciprocity of spin-wave dispersions in heterostructures consisting of two distinct ferromagnetic materials, focusing on the influence of saturation magnetization and thickness of the magnetic layers.
We exploit Brillouin light scattering to confirm numerical calculations which are conducted with the finite element software \textsc{TetraX}. An extensive numerical analysis reveals that the nonreciprocal behavior is strongly influenced by the changing material parameters, with asymmetry in the spin-wave propagation direction reaching several GHz under optimized conditions. Our findings demonstrate that tailoring the bilayer composition enables precise control over nonreciprocity, providing a pathway for engineering efficient unidirectional spin-wave devices. These results offer a deeper understanding of hybrid ferromagnetic systems and open avenues for designing advanced magnonic circuits.

\end{abstract}

\maketitle

%%%%%%%%%%%%%%%%%%%%%%%%%%%%%%%%%%%%%%
%%%%%%%%% Introduction
%%%%%%%%%%%%%%%%%%%%%%%%%%%%%%%%%%%%%%

\section*{Introduction}
Spin waves are the collective excitations in magnetically ordered materials. Their quanta are known as magnons with energies in the $\mu$eV range and frequencies about several GHz.
The history of spin waves dates back to the discovery of the ferromagnetic resonance which can be considered as a spin wave with infinite wavelength or zero wave vector. 
Already in the 1960s, it was discovered theoretically by Damon and Eshbach that spin waves restricted to finite geometries show nontrivial wave phenomena\cite{damonMagnetostaticModesFerromagnet1961}. They reported on magnetostatic surface waves (MSSW) which only exist in thin magnetic films with the magnetization lying in the film plane and for propagation in a narrow angular range around the direction perpendicular to the magnetization. 
One peculiarity of these MSSW is the fact that the mathematical equations only allow for a solution with a unidirectional propagation on one surface of the film~\cite{kostylevNonreciprocityDipoleexchangeSpin2013}. This implies that it is impossible to invert the propagation direction of the MSSW on the same surface unless the direction of the magnetization is reversed as well. 

Initially, this nonreciprocity of spin waves gained little attention. But with the discovery of magnetic skyrmions~\cite{bogdanovMagneticStructuresReorientation2002b} and the Dzyaloshinskii–Moriya interaction\cite{dzyaloshinsky_thermodynamic_1958,moriya_new_1960}, which introduces an asymmetric term in the spin Hamiltonian of magnetic materials, the interest in spin waves and their nonreciprocal behavior increased significantly.
In essence, it is the chiral nature of MSSW discussed by Damon and Eshbach which causes an additional frequency splitting of spin waves propagating in opposite directions. From this point on, spin waves became attractive as a sensitive and precise probe to quantify the strength of the asymmetric exchange interaction \cite{nembachLinearRelationHeisenberg2015}.
Furthermore, the discovery of asymmetric exchange and the resulting nonreciprocity of the spin-wave dispersion~\cite{zakeriAsymmetricSpinWaveDispersion2010,cortes-ortunoInfluenceDzyaloshinskiiMoriya2013} stimulated the field of magnonics, where spin waves are envisioned as carriers of information, promising low-loss and analogue computing features\cite{neusser_magnonics_2009,chumak_magnon_2015, chumak_advances_2022,flebus_2024_2024}.
Unfortunately, most attempts to exploit the interfacial Dzyaloshinskii–Moriya interaction or surface anisotropies in this respect failed because either the frequency splitting was only a few tens of MHz \cite{hrabecMakingDzyaloshinskiiMoriyaInteraction2017,zhang_quantifying_2022,khaliliamiriNonreciprocalSpinWave2007,haidarNonreciprocalOerstedField2014,gladiiFrequencyNonreciprocitySurface2016a} or the respective materials exhibited larger magnetic damping, prohibiting spin-wave propagation over distances more than a few micrometer\cite{quessabTuningInterfacialDzyaloshinskiiMoriya2020}.
In the meantime, it has been demonstrated theoretically that spin waves propagating on curved magnetic surfaces exhibit a distinguished interplay between the dynamic magnetic charges in the volume and on the surfaces causing a sizable frequency splitting of counter propagating spin waves in the range of several hundreds of MHz\cite{otaloraCurvatureInducedAsymmetricSpinWave2016,otaloraAsymmetricSpinwaveDispersion2017,salazar-cardonaNonreciprocitySpinWaves2021}. However, the required small radii of the ferromagnetic tubes made it difficult to realize devices. 
In subsequent works \cite{cho_thickness_2015, gallardoReconfigurableSpinWaveNonreciprocity2019, gallardoSpinwaveNonreciprocityMagnetizationgraded2019}, a strong nonreciprocity was found for simple magnetic bilayers~\cite{henryPropagatingSpinwaveNormal2016,grassiSlowWaveBasedNanomagnonicDiode2020}, exchange-coupled bilayers~\cite{slukaEmissionPropagation1D2019,gallardoReconfigurableSpinWaveNonreciprocity2019} or single layer films with a gradient of the magnetization along the film thickness~\cite{gallardoSpinwaveNonreciprocityMagnetizationgraded2019}.
In these systems, finally, the nonreciprocity was large enough to be quantified in a small range of wave vectors using propagating spin-wave spectroscopy, to build a prototype spin-wave diode \cite{grassiSlowWaveBasedNanomagnonicDiode2020} and demonstrate unidirectional propagation in several systems~\cite{shichiSpinWaveIsolator2015,thiancourtUnidirectionalSpinWaves2024,wojewodaUnidirectionalPropagationZeromomentum2024}.

In this work, we present a systematic experimental study of the spin-wave dispersion in magnetic bilayers to thoroughly test the results of the simulation package \textsc{TetraX}\cite{korberFiniteelementDynamicmatrixApproach2022,korberTetraXFiniteElementMicromagneticModeling2022} and subsequently use it to optimize the bilayer parameters for non reciprocal propagation. This recently developed open-source finite-element micromagnetic package provides a precise and efficient method for computing and analyzing spin-wave dispersions in magnetic thin-film heterostructures as well as arbitrary shaped infinitely long waveguides. The excellent agreement between experimental measurements and simulation results strengthens confidence in the model and enables its application in predicting spin-wave nonreciprocity in a wide range of bilayer geometries and material compositions. Furthermore, simulations of spin-wave mode profiles across the film thickness reveal the spatial localization of spin-wave energy as a function of frequency and wave vector. Additionally, our simulations allow us to extract the sense of magnetization precession across the film thickness. In thicker films, the hybridization of different spin-wave modes quantized along the film thickness leads to anti-Larmor precession of the magnetization at specific points along the film thickness.

%%%%%%%%%%%%%%%%%%%%%%%%%%%%%%%%%%%%%%
%%%%%%%%% FIGURE schematics
%%%%%%%%%%%%%%%%%%%%%%%%%%%%%%%%%%%%%%
    \begin{figure}
    \includegraphics{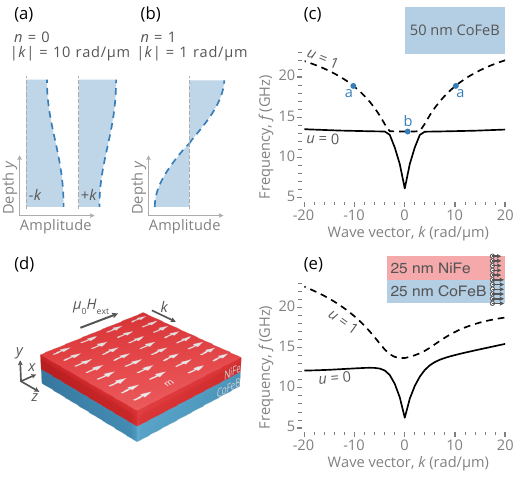}
    \caption{(a) Thickness profiles for MSSW with opposite wave vectors $k$ and $-k$ in a \SI{50}{\nano\meter} thick CoFeB film. (b) Thickness profiles for the PSSW mode with $n=1$ in a \SI{50}{\nano\meter} thick CoFeB film. (c) Spin-wave dispersion for a \SI{50}{\nano\meter} thick layer of CoFeB. (d) Sketch of a ferromagnetic bilayer with the magnetization aligned along an external magnetic field $B_{\text{ext}}$ and spin-wave propagation perpendicular to this. (e) Nonreciprocal dispersion for a multilayer with broken symmetry along the thickness. The inset depicts the magnetization parameters changing along the multilayer thickness as indicated in the one-dimensional line-trace mesh used in the \textsc{TetraX} simulations. }
    \label{fig:schematic}
    \end{figure}
%%%%%%%%%%%%%%%%%%%%%%%%%%%%%%%%%%%%%%

\section*{Methods}

The focus of our study lies on the nonreciprocity of the spin-wave dispersion upon reversal of the spin-wave wave vector ($k \to -k$), and how this nonreciprocity can be designed by choosing the appropriate materials and layer thicknesses.
As mentioned before, even in a single magnetic layer, MSSW exhibit an asymmetric mode profile with decaying amplitude along the film thickness. For opposite propagation direction, their amplitude is larger at opposing surfaces, as depicted in Fig.~\figref{fig:schematic}{a}.
Additionally, quantization across the film thickness becomes relevant for thicknesses of a few tens of nanometer, as studied in this work. Typically, these higher order perpendicular standing spin waves (PSSW) are characterized by their mode number $n$, counting the nodes across the film thickness (e.g. $n=1$ is plotted in Fig.~\figref{fig:schematic}{b}) and have frequencies in the range of the MSSW. Figure~\figref{fig:schematic}{c} shows the dispersion relation of the MSSW and the first-order PSSW mode for a \SI{50}{\nano\meter} thick, homogeneously magnetized CoFeB film. 
The crossing of the MSSW with the PSSW mode leads to their hybridization\cite{tacchiStronglyHybridizedDipoleexchange2019,Klingler2018-np}. That is why one spin-wave mode, which we plot by different line styles in Fig.~\figref{fig:schematic}{c}, changes its quantization character as a function of wave vector and $n$ is not a proper characterization any more. Hence, we introduce a different parameter $u=0,1,2,...$ to label the different spin-wave modes.
Despite the hybridization of the MSSW and the PSSW mode, the spin-wave frequencies are the same for $k$ and $-k$.

This changes drastically in a bilayer system such as schematically shown in Fig.~\figref{fig:schematic}{d}. Here, we introduce a bilayer of two metallic ferromagnets CoFeB and NiFe with significantly different saturation magnetization and exchange constant.
The magnetization direction is fixed along an external magnetic field $B_{\text{ext}}$. As before, we focus on the propagation perpendicular to the magnetization direction. In Fig.~\figref{fig:schematic}{e}, we plot the dispersion of the two lowest spin-wave modes $u=0,1$. Due to the strong asymmetry in the magnetic properties across the film thickness, spin waves propagating in opposite directions are influenced differently and, thereby, have different frequencies for $k$ compared to $-k$.
Additionally, the much stronger and wave-vector dependent hybridization of the different order thickness modes further increases the nonreciprocity of the spin-wave dispersion. 

\subsection*{Theory and simulation}

For the simulations in this work, the open-source finite element software \textsc{TetraX}\cite{korberTetraXFiniteElementMicromagneticModeling2022} is used which allows for a direct and fast numerical calculation of the spin-wave dispersion for the bilayer systems investigated here. The method does not rely on time integration of the Landau–Lifshitz–Gilbert (LLG) equation but instead allows for the direct calculation of the system's eigenmodes via a propagating-wave dynamic matrix approach\cite{henryPropagatingSpinwaveNormal2016,korberFiniteelementDynamicmatrixApproach2021}. In the case of thin films, as used in this work, it is sufficient to simulate only a one-dimensional line trace across the film thickness\cite{korberFiniteelementDynamicmatrixApproach2022}. Therefore, the calculation of the dispersion relation for one parameter set takes only seconds on a standard laptop. This makes it possible to systematically study  multiple parameters influencing the nonreciprocity. The line-trace mesh is depicted in the inset of Fig.~\figref{fig:schematic}{e} and has a resolution of \SI{1}{\nano\meter}.

Throughout this work, we keep the same parameters for all simulations involving Ni$_{81}$Fe$_{19}$ and Co$_{40}$Fe$_{40}$B$_{20}$. 
For Ni$_{81}$Fe$_{19}$, we use a saturation magnetization of $M_\mathrm{s}=\SI{830}{\kilo\ampere/\meter}$ and an exchange constant of
$A=\SI{13}{\pico\joule/\meter}$. The Co$_{40}$Fe$_{40}$B$_{20}$ was simulated using $M_\mathrm{s}=\SI{1260}{\kilo\ampere/\meter}$ and $A=\SI{16}{\pico\joule/\meter}$. A gyromagnetic ratio of $\gamma= \SI{176}{\radian\giga\hertz/\tesla}$ is assumed for both materials. In later simulations, the magnetization is systemically varied, which will be discussed in more detail for the respective figures.

%%%%%%%%%%%%%%%%%%%%%%%%%%%%%%%%%%%%%%
%%%%%%%%% FIGURE experiment
%%%%%%%%%%%%%%%%%%%%%%%%%%%%%%%%%%%%%%
    \begin{figure*}
    \includegraphics{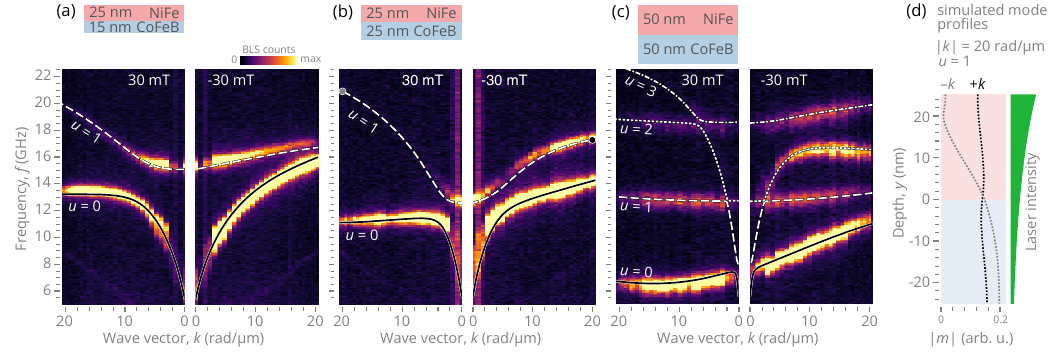}
    \caption{(a)-(c) Spin-wave dispersions of NiFe/CoFeB bilayer systems measured with conventional BLS and overlaid with simulated results from \textsc{TetraX}. The color map depicts the Stokes signal for the two magnetic field directions. In these bilayer systems, the mode profile is highly asymmetric. For the two marked wave vectors in (b), the mode profile is extracted from the simulation and shown in (d). This asymmetric profile, in combination with the limited penetration depth for the probing laser (d), results in a weaker signal for positive fields, where the mode is more localized to the CoFeB layer. }
    \label{fig:exp}
    \end{figure*}
%%%%%%%%%%%%%%%%%%%%%%%%%%%%%%%%%%%%%%

\subsection*{Brillouin light-scattering spectroscopy}

Spin-wave dispersion relations are measured using Brillouin light scattering (BLS) spectroscopy\cite{demokritovBrillouinLightScattering2001, sebastian_micro-focused_2015}. BLS is based on the inelastic scattering of photons with spin waves. Therefore, a single mode laser with a wavelength of  $\lambda=\SI{532}{\nano\meter}$ and a beam diameter of \SI{3}{\milli\meter} is focused onto the sample plane using a 2-inch achromatic lens with \SI{75}{\milli\meter} focal length. Thereby, we achieve a well-defined angle of incidence $\theta$ between the sample normal and the incoming laser. The laser power is set to \SI{200}{\milli\watt} at the sample. 
The backscattered light is directed to a Tandem-Fabry-Pérot interferometer (TFPI)\cite{mock_construction_1987} for spectral analysis. The frequency shift of the inelastically scattered light is analyzed to determine the spectrum of thermally activated magnons. The intensity of the scattered light is directly proportional to the spin-wave intensity. Due to momentum conservation, the measured wave vector $k$ can be selected by changing the angle of incidence $\theta$ via: $k=4\pi\lambda\sin{\theta}$.
To achieve this wave-vector selectivity, we mount the sample on a rotational stage inside an electromagnet with the axis of rotation parallel to the film plane and the magnetic field but perpendicular to the incident laser. The used electromagnet can provide a maximum field of \SI{1.5}{\tesla}. All measurements are performed at room temperature.

Three approaches can be used for quantifying the nonreciprocity of the spin-wave dispersion with BLS, each of which has its potential drawbacks: 
First, one can directly compare the negative (Stokes signal) and positive frequency shifts (anti-Stokes signal), corresponding to $-k$ and $+k$ respectively. However, even a slight misalignment between the laser beam probing the sample and the reference laser beam, which defines the zero-GHz line of the TFPI, can introduce frequency shifts between the Stokes and anti-Stokes signals, affecting the accuracy of the measurements. 
Second, measuring the spectra for positive and negative angles of incidence causes a shift of Stokes and anti-Stokes signals. However, this approach is susceptible to errors arising from sample misalignment relative to the angle of incidence.
Third, the direction of the magnetic field can be inverted to reverse the propagation direction of spin waves. This method is safe to apply if the magnitude of the magnetic field is sufficiently large to guarantee saturation of the magnetic film, and if the magnetic field is constantly measured by a magnetic field sensor to rule out any negative effects of a hysteretic behaviour of the magnet system. The frequency shift can be determined more accurately since the external magnetic field can be set more precisely. 
Hence, we refer to this third method and measure the spin-wave spectra for both field directions for each angle of incidence. 

\subsection*{Sample fabrication}

The magnetic materials are deposited on an intrinsic Si/Si$\text{O}_2$ substrate using dc magnetron sputtering from targets with the nominal composition $\text{Co}_{\text{40}}\text{Fe}_{\text{40}}\text{B}_{\text{20}}$ and $\text{Ni}_{\text{81}}\text{Fe}_{\text{19}}$. Before and after depositing the magnetic bilayer, we sputter \SI{5}{\nano\meter} and \SI{3}{\nano\meter} thick Cr layers to promote the growth and prevent oxidation, respectively. The thicknesses $t_{\text{NiFe}}$ and $t_{\text{CoFeB}}$ of the magnetic materials are varied between \SI{15}{\nano\meter} and \SI{50}{\nano\meter}, resulting in total film thicknesses up to $t_\text{total}=\SI{100}{\nano\meter}$.

\section*{Results}

\subsection*{Experimental verification of the simulation model}

Figure~\figref{fig:exp}{a-c} shows spin-wave dispersion relations of three different layer combinations with the experimentally acquired BLS data color-coded.
Note that the $x$-axes left and right from $k=0$ cover the same range (from 0 to \SI{20}{\radian/\micro\meter}) for opposite applied fields $\mu_0H=\pm\SI{30}{\milli\tesla}$ but the left part is plotted inverted going from large to small values. We have chosen this presentation to ease comparison with spin-wave dispersion relations plotted from negative to positive wave vectors for one magnetic field direction.

For the thinnest bilayer film with $t_\text{CoFeB}=\SI{15}{\nano\meter}$ and $ t_\text{NiFe}=\SI{25}{\nano\meter}$, we measure two spin-wave modes, with the lower one ($u=0$) rapidly increasing for small wave vectors resembling the character of MSSW. For larger wave vectors, however, the two spin-wave modes obviously differ from the dispersion of a single-layer film due to the strong hybridization of the MSSW with the first-order PSSW. This results in a strong nonreciprocity for counterpropagating spin waves in both modes of the dispersion. 

As the total thickness increases in Fig.~\figref{fig:exp}{b}, the frequency of the PSSW mode decreases, which leads to a more pronounced frequency nonreciprocity. For the thickest bilayer film with $t_\text{total}=\SI{100}{\nano\meter}$ in Fig.~\figref{fig:exp}{c}, even more higher order modes enter the investigated frequency range, resulting in a much more complex dispersion.
The additional modes are flatter and show less nonreciprocity. Furthermore, the hybridization of the MSSW mode and the first PSSW mode, whose frequency is close to the ferromagnetic resonance, results in a negative group velocity for the positive field direction. 

Using \textsc{TetraX}, we simulate the first four eigenmodes for an external field of $\mu_0H\text{ext}=\SI{30}{\milli\tesla}$ with a resolution of \SI{0.25}{\radian/\micro\meter}. The results are plotted as lines on top of the BLS data with different line styles referring to different spin-wave modes $u=0,1,2,3$. While in the measurement the propagation direction reversal was achieved by switching the field direction, the simulated eigenmodes are calculated with a fixed external field for positive and negative wave vector $\pm k$. An excellent agreement between the simulated and the experimentally acquired dispersion is seen for all three films, validating the simulations.

The main difference between the simulated and measured dispersion is the absence of modes in the experimental data. For example, in Fig.~\figref{fig:exp}{b}, the mode $u=2$ at $k=\SI{-3}{\radian/\micro\meter}$ marked with a grey dot in the positive field direction does not appear in the measurement. 
This is the result of the limited penetration depth of the laser. The signal is acquired mainly at the top NiFe surface. However, the out-of-plane intensity of the spin-wave modes is primarily located in the CoFeB layer for one propagation direction, as shown in Fig.~\figref{fig:exp}{d}. Therefore, its detection is much less efficient with laser spectroscopy. To support this, the laser penetration depth is shown next to the spin-wave intensity profiles in Fig.~\figref{fig:exp}{d}. It is calculated according to the formula with the simplified assumption of a single NiFe film:
\begin{equation}
    \delta_{\text{penetration}}=\frac{c}{\omega\Im{\epsilon_{xx}}}=\SI{26}{\nano\meter}
\end{equation}
with $c$ the speed of light, $\omega$ the laser frequency, and $\Im(\epsilon_{xx})$ the imaginary part of the diagonal permittivity tensor component for NiFe taken from Tikui\v{s}is et al.\cite{tikuivsis2017optical}. This confirms that the BLS is mainly sensitive to the top NiFe layer, especially for larger total thicknesses.

%%%%%%%%%%%%%%%%%%%%%%%%%%%%%%%%%%%%%%
%%%%%%%%% FIGURE field
%%%%%%%%%%%%%%%%%%%%%%%%%%%%%%%%%%%%%%
    \begin{figure}
    \includegraphics{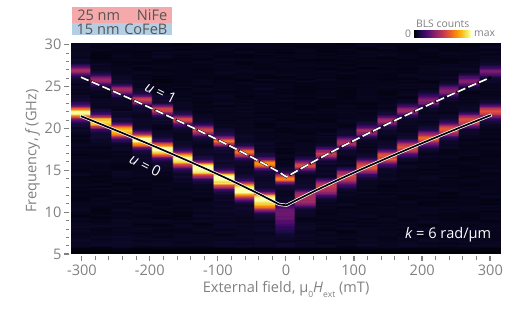}
    \caption{External-field dependence of the two spin-wave modes with $k=\SI{6}{\radian/\micro\meter}$ measured with BLS at a fixed angle of incidence on the thinnest bilayer with $t_{\text{CoFeB}}=\SI{15}{\nano\meter}$ and $ t_{\text{NiFe}}=\SI{25}{\nano\meter}$. Lines show the results from \textsc{TetraX} simulations.}
    \label{fig:field}
    \end{figure}
%%%%%%%%%%%%%%%%%%%%%%%%%%%%%%%%%%%%%%

To further verify the micromagnetic simulations, we take the same film as in Fig.~\figref{fig:exp}{a} with $t_{\text{CoFeB}}=\SI{15}{\nano\meter}$ and $ t_{\text{NiFe}}=\SI{25}{\nano\meter}$ and measure the thermal BLS spectra at a fixed angle as a function of the external magnetic field.
We chose a wave vector of $k=\SI{6}{\radian/\micro\meter}$ to get a strong signal for the first two spin-wave modes ($u=0,1$). In the simulation, the external field was changed in $\SI{10}{\milli\tesla}$ steps. At each field, the magnetization was relaxed. The minimum external field was set to $\mu_0H_0=\SI{1}{\milli\tesla}$. As shown in Fig.~\ref{fig:field}, the experimental results and \textsc{TetraX} simulations agree well, even for the extended field range.

\subsection*{Systematic variation of the magnetization}

%%%%%%%%%%%%%%%%%%%%%%%%%%%%%%%%%%%%%%
%%%%%%%%% FIGURE magnetization
%%%%%%%%%%%%%%%%%%%%%%%%%%%%%%%%%%%%%%
    \begin{figure}
    \includegraphics{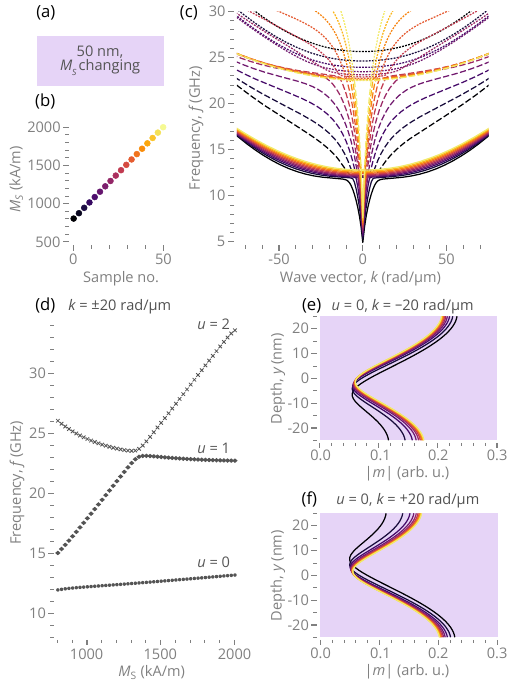}
    \caption{(a) \SI{50}{\nano\meter} thick, homogeneously magnetized single layer with varying magnetization color-coded as shown in (b). (c) Dispersion relation for the various magnetization values. (d) Frequency of the spin-wave modes $u=0,1,2$, extracted at a wave vector $\pm k=\SI{20}{\radian/\micro\meter}$ as a function of $M_\text{S}$. Due to the reciprocal dispersion, $+k$ and $-k$ have the same frequencies. (e),(f) Comparison between mode profiles of the lowest mode $u=0$ for counter-propagating waves with $\mathopen|k\mathclose|=\SI{20}{\radian/\micro\meter}$ and the value of $M_\text{S}$ color coded as in (b).}
    \label{fig:magnetisation_supp}
    \end{figure}
%%%%%%%%%%%%%%%%%%%%%%%%%%%%%%%%%%%%%%

With the numerical implementation confirmed by experimental results, we now focus on \textsc{TetraX} simulations to study the influence of changing material parameters on the dispersion relation in more detail. As mentioned before, the nonreciprocity depends on various parameters, ranging from the film thickness over the saturation magnetization to the exchange constant. In this work, we focus on the influence of the saturation magnetization and layer thickness. 

For reference, we begin simulating a \SI{50}{\nano\meter} thick, single layer (Fig.~\figref{fig:magnetisation_supp}{a}) and continuously change the magnetization therein. As depicted in Fig.~\figref{fig:magnetisation_supp}{b}, we go from a pure NiFe film with $M_\mathrm{s}=\SI{800}{\kilo\ampere/\meter}$ to a film with a much higher saturation magnetization $M_\mathrm{s}=\SI{2000}{\kilo\ampere/\meter}$, keeping the exchange constant fixed at $A_{\text{ex}}=\SI{13}{\pico\joule/\meter}$. We simulate dispersion relations for 50 different saturation magnetization, 18 of which are summarized in Fig.~\figref{fig:magnetisation_supp}{c}, with the value of $M_\mathrm{s}$ color coded as in Fig.~\figref{fig:magnetisation_supp}{b}. As expected, the dispersion relations are reciprocal and the frequencies generally increase with increasing $M_\mathrm{s}$, however, not homogeneously across all spin-wave modes $u$ and all wave vector $k$. 
To analyze this more quantitatively, Fig.~\figref{fig:magnetisation_supp}{d} shows how the frequencies of the different modes evolve for a fixed wave vector $k=\pm\SI{20}{\radian/\micro\meter}$ when gradually increasing the saturation magnetization.
While the frequency of the lowest spin-wave mode $u=0$ only changes slightly, the frequencies of modes $u=1,2$ increase more rapidly. Interestingly, when mode $u=1$ reaches the lowest frequency value of mode $u=2$ across all $M_\mathrm{s}$, the frequency in $u=1$ does not rise anymore even though $M_\mathrm{s}$ is still increasing.

In Fig.~\figref{fig:magnetisation_supp}{e, f}, we show the mode profiles for the lowest spin-wave mode $u=0$ for $k=\pm\SI{20}{\radian/\micro\meter}$ by plotting the magnitude $|m|$ of the magnetization vector over the depth $y$:
\begin{equation}
    \begin{split}
    |m|
    &=\Big(\Re{m_x}^2+\Im{m_x}^2\\
    &+\Re{m_y}^2+\Im{m_y}^2\\
    &+\Re{m_z}^2+\Im{m_z}^2\Big) ^{1/2}.
    \end{split}
\end{equation}

Due to the hybridization of the MSSW and the first-order PSSW mode, the magnitude never decreases to zero, even in the simple single layer. However, the surface character is still visible in the sense that opposite wave vectors show a concentration of the spin-wave magnitude on opposite surfaces of the film. 

%%%%%%%%%%%%%%%%%%%%%%%%%%%%%%%%%%%%%%
%%%%%%%%% FIGURE magnetization
%%%%%%%%%%%%%%%%%%%%%%%%%%%%%%%%%%%%%%
    \begin{figure}
    \includegraphics{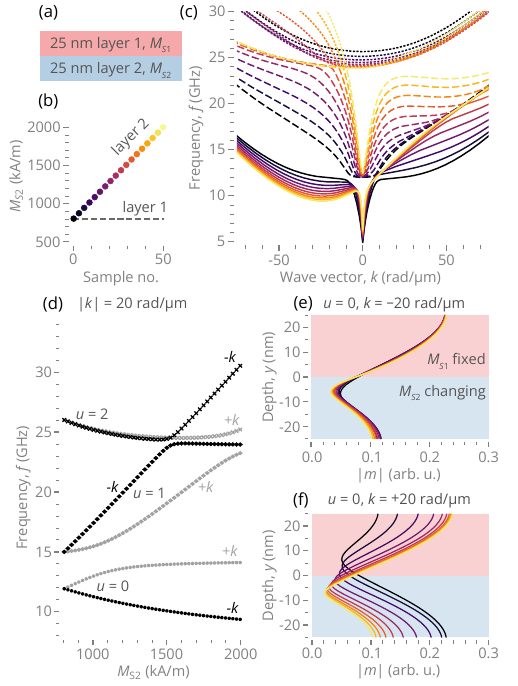}
    \caption{(a) The thicknesses of both layers are constant, with (\SI{25}{\nano\meter}) each. (b) Layer 1 has a fixed $M_\text{S1}=\SI{800}{\kilo\ampere/\meter}$ corresponding to NiFe. In layer 2, the saturation magnetization changes continuously from \SI{800}{\kilo\ampere/\meter} to \SI{2000}{\kilo\ampere/\meter}. (c) Spin-wave dispersion relations for varying saturation magnetization of layer 2, with the values of $M_\mathrm{s}$ color coded as in (b). (d) Frequency of the different spin-wave modes $u=0,1,2$ extracted for the wave vector $\mathopen|k\mathclose|=\SI{20}{\radian/\micro\meter}$ showing their difference for opposing propagation directions. (e),(f) Comparison between mode profiles of the lowest mode $u=0$ for counter-propagating waves with $\mathopen|k\mathclose|=\SI{20}{\radian/\micro\meter}$ and the value of $M_\text{S}$ color-coded as in (b).}
    \label{fig:magnetisation}
    \end{figure}
%%%%%%%%%%%%%%%%%%%%%%%%%%%%%%%%%%%%%%

%%%%%%%%%%%%%%%%%%%%%%%%%%%%%%%%%%%%%%
%%%%%%%%% FIGURE layer
%%%%%%%%%%%%%%%%%%%%%%%%%%%%%%%%%%%%%%
    \begin{figure*}
    \includegraphics{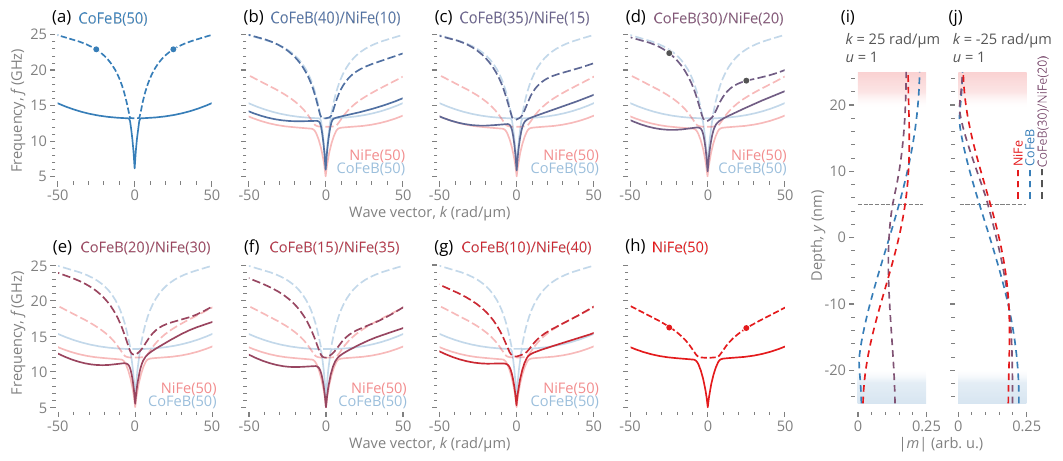}
    \caption{Spin-wave dispersions of a \SI{50}{\nano\meter} thick film for changing ratio between CoFeB and NiFe showing the transition from (a) a pure CoFeB film to (h) a pure NiFe film. (b)-(g) show asymmetric dispersions for the intermediate bilayer systems. (i),(j) Comparison of mode profiles extracted for the spin-wave mode $u=1$ at $|k|=\SI{25}{\radian/\micro\meter}$ for a pure NiFe film (red), a pure CoFeB film (blue) and a bilayer with $t_\text{CoFeB}=\SI{30}{\nano\meter}$ and  $t_\text{NiFe}=\SI{20}{\nano\meter}$. The red (blue) shade indicates that the NiFe (CoFeB) is at the top (bottom) surface of the bilayer. }
    \label{fig:layer}
    \end{figure*}
%%%%%%%%%%%%%%%%%%%%%%%%%%%%%%%%%%%%%%

Now we move to a bilayer with equal thicknesses \SI{25}{\nano\meter}/\SI{25}{\nano\meter}, as sketched in Fig.~\figref{fig:magnetisation}{a}, but changing saturation magnetization in one of the layers (layer 2), as color coded in Fig.~\figref{fig:magnetisation}{b}. Starting point is a pure NiFe film with  \SI{50}{\nano\meter} thickness, $M_\mathrm{s}=\SI{800}{\kilo\ampere/\meter}$ and $A_{\text{ex}}=\SI{13}{\pico\joule/\meter}$. Then, the saturation magnetization in half of the heterostructure is continuously increased from \SIrange{800}{2000}{\kilo\ampere/\meter} while keeping the exchange constant fixed.
Figure~\figref{fig:magnetisation}{c} summarizes the different dispersion relations with the value of $M_\mathrm{s}$ color coded as in Fig.~\figref{fig:magnetisation}{b}. With a more pronounced difference in the magnetization, the positive and negative wave-vector sides evolve in opposite directions for the lowest frequency mode $u=0$. For positive wave vectors, the frequencies of mode $u=0$ continuously increase until the second mode ($u=1$) of a homogeneous NiFe film is approached. At this point, the frequency does not increase further even though the saturation magnetization still increases. This means that there is a maximum frequency which is not exceeded.
In contrast, for negative wave vectors, the frequencies of the lowest mode $u=0$  continuously decrease, opening the gap to the $u=1$ mode. This is highlighted in Fig.~\figref{fig:magnetisation}{d}, where we extract the frequencies of spin-wave modes $u=0,1,2$ at a fixed wave vector
$k=\pm\SI{20}{\radian/\micro\meter}$.

The frequencies of the second spin-wave mode $u=1$ increase for both propagation directions. The change of $f_{u=1}$ for negative wave vectors resembles the behaviour in the homogeneous film [Fig.~\figref{fig:magnetisation_supp}{d}] with the frequency saturating when it reaches the minimum frequency of the next order mode $u=2$. 
The difference introduced by the bilayer lies in the positive wave-vectors, for which the frequencies increase much slower.

For $\mathopen|k\mathclose|=\SI{20}{\radian/\micro\meter}$ and $u=0$, the mode profiles are shown in Fig.~\figref{fig:magnetisation}{e,f}. The modes $u=0$ and $u=1$ hybridize at this wave vector but keep their surface character. As expected for MSSW, the magnitude of a homogeneous film [black line in Fig.~\figref{fig:magnetisation}{e,f}] has a stronger localization at one surface which is mirrored under the reversal of the propagation direction to the other surface of the film. 
When the magnetization is increased in half of the film, the mode profiles do not change significantly for negative wave vectors. For positive $k$, however, the mode localization shifts from the one surface with fixed $M_\mathrm{s}$ to the other with increasing $M_\mathrm{s}$. This means the nonreciprocal character of the surface localization is lost. The hybridization between the first PSSW and the MSSW mode will be discussed in more detail later in the manuscript.

%%%%%%%%%%%%%%%%%%%%%%%%%%%%%%%%%%%%%%
%%%%%%%%% FIGURE delta f 50
%%%%%%%%%%%%%%%%%%%%%%%%%%%%%%%%%%%%%%
    \begin{figure}
    \includegraphics{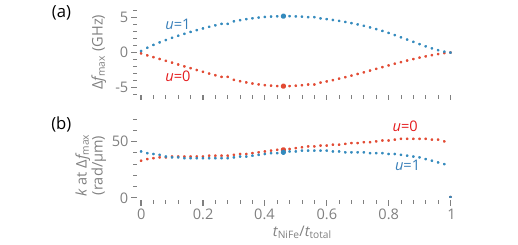}
    \caption{(a) Largest frequency difference for two counter-propagating waves of the same wavelength in a \SI{50}{\nano\meter} thick bilayer of varying CoFeB/NiFe ratio. Red (blue) depicts the lowest (second to lowest) frequency mode. (b) Wave vector for the largest frequency difference for the different layer combinations. The dot size is increased for the thickness of the largest nonreciprocity.}
    \label{fig:deltaf}
    \end{figure}
%%%%%%%%%%%%%%%%%%%%%%%%%%%%%%%%%%%%%%

%%%%%%%%%%%%%%%%%%%%%%%%%%%%%%%%%%%%%%
%%%%%%%%% FIGURE different thicknesses
%%%%%%%%%%%%%%%%%%%%%%%%%%%%%%%%%%%%%%
    \begin{figure*}
    \includegraphics{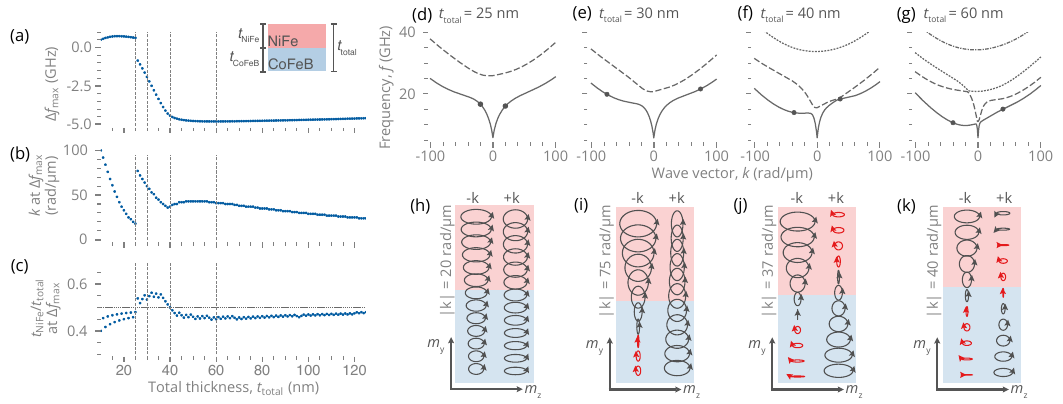}
    \caption{Analysis of the maximal nonreciprocity for NiFe/CoFeB bilayers with different total thicknesses $t_\text{total}$. (a) Maximum frequency difference for two counter-propagating, lowest-order spin-wave modes ($u=0$) with the same wave vector $k$. (b) Wave vector and (c) proportion of the NiFe layer in percent of the total layer thickness for which $\Delta f_\text{max}$ is reached. (d)-(g) Dispersion relations calculated for the indicated total thicknesses and layer combinations as shown in (c). Wave vectors yielding the strongest nonreciprocity are indicated by dots. (h)-(k) Mode profiles for the wave vectors yielding the strongest nonreciprocity. The left (right) profile in each panel corresponds to negative (positive) wave vectors, respectively. Trajectories plotted in red are the areas with reversed precession, corresponding to the onset of hybridization between the first two spin-wave modes.}
    \label{fig:thick}
    \end{figure*}
%%%%%%%%%%%%%%%%%%%%%%%%%%%%%%%%%%%%%%

%layer thickness
\subsection*{Changing layer thickness}

In the next step of our analysis, we fix the total layer thickness to $t_\text{total}=\SI{50}{\nano\meter}$. To study the transition from a pure CoFeB film to a pure NiFe film, we continuously change the ratio of the two thicknesses $t_\text{NiFe}$ and $t_\text{CoFeB}$.
The material parameters are the same as in Fig.~\ref{fig:exp} with 
$\mu_0M_s=\SI{830}{\kilo\ampere/\meter}$,
$A_{ex}=\SI{13}{\pico\joule/\meter}$ for NiFe and $\mu_0M_s=\SI{1260}{\kilo\ampere/\meter}$,
$A_\text{ex}=\SI{16}{\pico\joule/\meter}$ for CoFeB. An external magnetic field of $\mu_0H_\text{ext}=\SI{30}{\milli\tesla}$ perpendicular to the propagation direction is applied for all simulations.

Figure~\ref{fig:layer} shows the transition from a pure CoFeB film [Fig.~\figref{fig:layer}{a}]to a pure NiFe film [Fig.~\figref{fig:layer}{h}]. For the mixed layer systems [Fig.~\figref{fig:layer}{b-g}], the homogeneous dispersion relations of NiFe and CoFeB are plotted as well in their respective colors.

Comparing the dispersion relations, the $u=0$ mode shows similar behavior as in the previous section for changing the magnetization. For one propagation direction ($+k$), the frequencies increase, whereas they decrease for the opposite direction ($-k$), even below that of either single-layer system. 
In contrast, the frequencies of the $u=1$ mode stay between the modes for the pure NiFe and CoFeB films. 

Due to the hybridization, the mode profile of the $u=1$ mode at a wave vector $\mathopen|k\mathclose|=\SI{25}{\radian/\micro\meter}$ resembles a typical mode profile of a MSSW, without any nodes across the film thickness for the pure NiFe film and only one node close to the surface of the pure CoFeB film [Fig.~\figref{fig:layer}{i,j}]. 
The mode profile for positive wave vectors has the majority of its energy in the upper layer, i.e. the NiFe. As a result, the dispersion for the $+k$ side of the $u=1$ mode resembles the single NiFe dispersion more closely. On the other hand, the negative wave vector is more localized at the bottom layer, i.e. the CoFeB. Therefore, the $-k$ dispersion of the heterostructures follows more closely the dispersion of the pure CoFeB layer, even for a balanced ratio of both magnetic materials, as seen in Fig.~\figref{fig:layer}{d}.

From Fig.~\ref{fig:layer}, it can be seen that heterostructures consisting of two materials with similar thicknesses exhibit the strongest nonreciprocity. To quantify this in more detail, we extract the maximum frequency difference $\Delta f=\mathopen|f(k)-f(-k)\mathclose|$ for two counter-propagating waves with wave vector $|k|$ from the dispersions calculated for all compositions with $t_\text{total}=\SI{50}{\nano\meter}$.
As is shown in Fig.~\ref{fig:deltaf}, the bilayer with $t_\text{NiFe}=\SI{23}{\nano\meter}$ and $ t_\text{CoFeB}=\SI{27}{\nano\meter}$ exhibits the largest frequency difference of $\Delta f_\text{max}=\SI{5.2}{\giga\hertz}$ for the $u = 1$ mode at a wave vector of $k_\text{max}=\SI{40.5}{\radian/\micro\meter}$. For the lowest spin-wave mode $u = 0$, the strongest nonreciprocity has an opposite sign and amounts to  $f_\text{max}=\SI{-4.8}{\giga\hertz}$ at a wave vector of $k_\text{max}=\SI{42.75}{\radian/\micro\meter}$.

To investigate if the nonreciprocity increases even more in overall thicker heterostructures, we repeat the previous simulation for NiFe/CoFeB bilayers with various total thicknesses $t_\text{total}$ ranging from \SIrange{10}{125}{\nano\meter}. For each total thickness, the percentage of the NiFe (CoFeB) layer is increased (decreased), respectively. From all these simulations, the largest frequency difference $\Delta f=\mathopen|f(k)-f(-k)\mathclose|$ for the spin-wave mode $u=0$ is extracted together with the corresponding wave vector $|k|$ and the percentage of the NiFe layer. Since for thinner films, the most significant differences occur at larger wave vectors, the range of the \textsc{TetraX} simulations is extended to \SI{100}{\radian/\micro\meter}.

Figure~\figref{fig:thick}{a-c} summarizes the results of this analysis. Initially, up to $t_\text{total}=\SI{25}{\nano\meter}$, the largest frequency difference of the $u=0$ mode is positive but relatively small. Then its sign changes, and $|\Delta f_\text{max}|$ increases drastically to a frequency difference of almost \SI{5}{\giga\hertz}. For $t_\text{total}>\SI{40}{\nano\meter}$, the frequency difference stays nearly constant, as does the layer ratio for which $\Delta f_\text{max}$ is reached, which is slightly below 50\%. Only the wave vector still changes to smaller values. 
This explains why the nonreciprocity in the measured data (Fig.~\ref{fig:exp}) is largest for the thickest film with $t_\text{total}=\SI{100}{\nano\meter}$. In BLS measurements, the accessible maximum wave vector is limited due to the limited momentum transfer. As the wave vector for maximum frequency difference decreases with increasing $t_\text{total}$, the nonreciprocity becomes more apparent in the measured range. For the thinner films, the maximum nonreciprocity occurs at wave vectors beyond the measurement range. 

As is exemplarily shown in Fig.~\figref{fig:thick}{d} for the bilayer with $t_\text{total}=\SI{25}{\nano\meter}$, no hybridization between the two lowest spin-wave modes is present in thin films with the small positive frequency differences. This is confirmed by the mode profiles, which are plotted in Fig.~\figref{fig:thick}{d} for $|k|=\SI{-20}{\radian/\micro\meter}$. Different from previous mode profiles, we plot the mode's trajectory in the $m_y$-$m_z$ plane, which is perpendicular to the applied field. As is expected without hybridization, the mode profile for $-k$ resembles a typical MSSW.
For $+k$, the mode profile is shifted into the layer with the larger saturation magnetization. The ellipticity in the two propagation directions is similar.

As the frequency difference reverses its sign for $\SI{25}{\nano\meter}<t_\text{total}< \SI{40}{\nano\meter}$, the lowest spin wave mode starts to hybridize with the first PSSW mode, which can be seen in Fig.~\figref{fig:thick} {i}. Here, the ellipticity of the mode changes significantly, and some areas even show a reversal of the precession direction, as marked in red.
Due to the different magnetization in the two layers, the onset of the hybridization is different for the opposing propagation directions. 

For $t_\text{total} \geq\SI{40}{\nano\meter}$, both modes are hybridized, as can be seen in Fig.~\figref{fig:thick}{j,k}. In contrast to the DE mode in Fig.~\figref{fig:thick}{h}, the larger intensity for these hybridized modes is not shifted towards the NiFe layer.
In addition, the precession becomes more flat in the layers with lower intensity. It should be noted that the occurrence of the hybridized mode with a locally reversed precession direction or anti-Larmor precession is not a consequence of the bilayer system. It has already been observed in single-layer films and was described as a hetero-symmetric spin-wave in previous works \cite{wintzHeterosymmetric, trevillian2024formation}.
Our simulations show that in a bilayer, this family of hybridized modes shows a negative frequency shift for the lower frequency and a positive frequency shift for the higher frequency mode. The frequency difference remains constant as the modes for both propagation directions show the hetero-symmetric character.
It is not further increasing even though the dipolar contribution should be larger for thicker films. 

%%%%%%%%%%%%%%%%%%%%%%%%%%%%%%%%%%%%%%
%%%%%%%%% FIGURE reversed precession
%%%%%%%%%%%%%%%%%%%%%%%%%%%%%%%%%%%%%%
    \begin{figure}
    \includegraphics{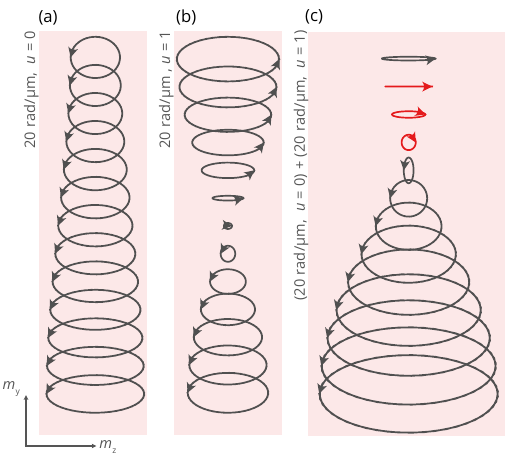}
    \caption{Origin of anti-Larmor precession. Precession of the (a) MSSW mode and (b) PSSW mode for a \SI{25}{\nano\meter} thick CoFeB layer at \SI{30}{\milli\tesla} external field. The arrowheads indicate the precession direction and its location shows the phase. (c) Superposition of the two lowest spin-wave modes under the assumption of a similar frequency, which results in one cell with flat precession and some areas with reversed precession direction.}
    \label{fig:reversal}
    \end{figure}
%%%%%%%%%%%%%%%%%%%%%%%%%%%%%%%%%%%%%%

While the higher-order PSSW modes remain largely unaffected by an inversion of the propagation direction [see Fig.~\figref{fig:thick}{f,g}], the $t_\text{total}$-dependent simulations further emphasize that the nonreciprocity is primarily based on the MSSW mode and its hybridization with the first PSSW mode.  
Qualitatively, this hybridized mode profile can be understood as a superposition of a MSSW mode and a PSSW mode at the same frequency. 
To illustrate this,
Fig.~\ref{fig:reversal} shows the mode profile for a \SI{25}{\nano\meter} thick CoFeB film. The MSSW mode [Fig.~\figref{fig:reversal}{a}] has a constant phase which is indicated by the arrowheads showing the position of the precession at a fixed time. The PSSW mode [Fig.~\figref{fig:reversal}{b}], however, shows a phase jump close to the middle of the film. Due to this jump, a phase difference occurs between the two modes in the upper half of the film. This, combined with the difference in ellipticity, results in the flat and partially reversed precession direction of the hybridized mode. 

\begin{figure}
    \includegraphics{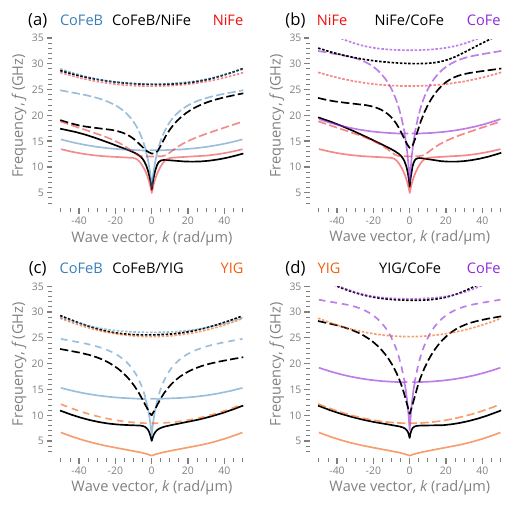}
    \caption{(a) Gray lines show the dispersion for a CoFeB/NiFe-bilayer system with the two materials having equal thicknesses of \SI{25}{\nano\meter}. The colored lines depict the dispersions for the \SI{50}{\nano\meter} thick homogeneous films of the respective material. (b) Dispersion of the same system, but the CoFeB is replaced by CoFe, a material with higher $M_\text{S}$, which further increases the nonreciprocity. (c),(d) Dispersions of material combinations including YIG, a material with low magnetization, which results in a much weaker nonreciprocity.}
    \label{fig:materials}
    \end{figure}

In the $t_\text{total}$-dependent simulation, it can be seen that the hetero-symmetric mode experiences the largest frequency nonreciprocity upon inversion of the propagation direction. However, this frequency difference remains steady around a maximum value when a critical film thickness is reached. To further increase the nonreciprocity, one material has to be replaced to increase the difference in saturation magnetization. In Fig.~\ref{fig:materials}, we plot dispersions for symmetric bilayer systems with different material combinations, all having a total thickness of $t_\text{total}=\SI{50}{\nano\meter}$.  

For direct comparison, Fig.~\figref{fig:materials}{a} shows again the dispersion for a symmetric CoFeB/NiFe bilayer with $t_\text{total}=\SI{50}{\nano\meter}$. In Fig.~\figref{fig:materials}{b} the CoFeB is replaced with CoFe ($M_\text{S}=\SI{1700}{\kilo\ampere/\meter}$, $A=\SI{26}{\pico\joule/\meter}$) showing a larger splitting. In contrast, replacing the NiFe with YIG ($M_\text{S}=\SI{140}{\kilo\ampere/\meter}$, $A=\SI{3.5}{\pico\joule/\meter}$), another material often used for magnonics, results in the opposite. As can be seen in Fig.~\figref{fig:materials}{c,d}, the nonreciprocity is less pronounced, even though the magnetization difference is larger. Here, the difference in material parameters is too large and the nonreciprocity remains small for the chosen symmetric layer stack.

%%%%%%%%%%%%%%%%%%%%%%%%%%%%%%%%%%%%%%
%%%%%%%%% FIGURE common materials
%%%%%%%%%%%%%%%%%%%%%%%%%%%%%%%%%%%%%%
%%%%%%%%%%%%%%%%%%%%%%%%%%%%%%%%%%%%%%

\section*{Discussion}

This study aimed to exploring the nonreciprocity of the spin-wave dispersion in heterostructures composed of two distinct
ferromagnetic materials with in-plane magnetization. We used the comparison of experimentally quantified dispersion relations for three different bilayer systems to verify the accuracy and robustness of the finite-element package \textsc{TetraX}. We found an excellent agreement between experiments and numerical calculations, which allowed us to use the numerical simulations for quantifying the frequency nonreciprocity, spin-wave depth profiles, and the handedness of the magnetization precession. 
First, we analyzed the spin-wave properties in films with a given total thickness when varying the saturation magnetization in one-half of the film. Surprisingly, after a first strong increase in the frequency splitting for counterpropagating waves, it saturates for the higher modes at around \SI{1500}{\kilo\ampere/\meter}, and, also the mode profiles do not shift anymore.
In the following, we kept the total film thickness constant at \SI{50}{\nano\meter} and shifted the ratio between the NiFe and CoFeB thicknesses. We found that the maximum nonreciprocity occurs when both layers have approximately the same thickness $t_\text{NiFe}=\SI{23}{\nano\meter}$ $t_\text{CoFeB}=\SI{27}{\nano\meter}$. The corresponding wave vector for a total film thickness of \SI{50}{\nano\meter} was around \SI{40}{\radian/\micro\meter}, already twice higher than what is accessible in typical $k$-resolved BLS experiments. 
Finally, we analyzed the impact of the total sample thickness on the nonreciprocity. The main consequence of increased total sample thickness is that the frequencies of the PSSW modes decrease such that the lowest PSSW mode strongly hybridizes with the MSSW. Once this happens, the lowest frequency modes for both propagation directions have a node in their thickness profile and show regions with anti-Larmor precession, where the magnetization is rotating opposite to the direction expected from simple Landau-Lifshitz torques.

We want to note that adjusting the saturation magnetization and the relative layer thicknesses is only a narrow section of the parameter space, which could be explored for designing spin-wave waveguides with intriguing properties. The impact of different exchange interaction, anisotropies, or even Dzyaloshinskii–Moriya interaction is still to be investigated.

%%%%%%%%%%%%%%%%%%%%%%%%%%%%%%%%%%%%%%
%%%%%%%%% Author contributions
%%%%%%%%%%%%%%%%%%%%%%%%%%%%%%%%%%%%%%

\section*{Author declarations}

\subsection*{Conflict of Interest}

The authors have no conflicts of interest to disclose.

\subsection*{Author contributions}

H.S. and O.G. conceptualized the presented work. 
H.S., O.G. and A.K. acquired funding.
A.K. and L.K. developed the code used for the numerical calculations. 
C.H. carried out the experiments, analyzed the data and performed the simulations.
V.I. fabricated the samples. All authors discussed the results.
C.H. and K.S. visualized the results.
C.H., H.S., A.K., K.S. wrote the original draft of the paper. 
All authors reviewed and edited the paper. 

%%%%%%%%%%%%%%%%%%%%%%%%%%%%%%%%%%%%%%
%%%%%%%%% Acknowledgements
%%%%%%%%%%%%%%%%%%%%%%%%%%%%%%%%%%%%%%

\section*{Acknowledgements}

This work was supported by the Deutsche Forschungsgemeinschaft (DFG) through the programs GL 1041/1-1 and KA 5069/3-1. 

\section*{Data availability}

The experimental BLS data as well as the script files used to generate the simulation data that support our findings are openly available in RODARE \cite{heins_christopher_2024_3336} and can readily be used with the open source software \textsc{TetraX}.

%%%%%%%%%%%%%%%%%%%%%%%%%%%%%%%%%%%%%%
%%%%%%%%% References
%%%%%%%%%%%%%%%%%%%%%%%%%%%%%%%%%%%%%%

\bibliography{references.bib,additional_references.bib}

\begin{thebibliography}{42}%
\makeatletter
\providecommand \@ifxundefined [1]{%
 \@ifx{#1\undefined}
}%
\providecommand \@ifnum [1]{%
 \ifnum #1\expandafter \@firstoftwo
 \else \expandafter \@secondoftwo
 \fi
}%
\providecommand \@ifx [1]{%
 \ifx #1\expandafter \@firstoftwo
 \else \expandafter \@secondoftwo
 \fi
}%
\providecommand \natexlab [1]{#1}%
\providecommand \enquote  [1]{``#1''}%
\providecommand \bibnamefont  [1]{#1}%
\providecommand \bibfnamefont [1]{#1}%
\providecommand \citenamefont [1]{#1}%
\providecommand \href@noop [0]{\@secondoftwo}%
\providecommand \href [0]{\begingroup \@sanitize@url \@href}%
\providecommand \@href[1]{\@@startlink{#1}\@@href}%
\providecommand \@@href[1]{\endgroup#1\@@endlink}%
\providecommand \@sanitize@url [0]{\catcode `\\12\catcode `\$12\catcode
  `\&12\catcode `\#12\catcode `\^12\catcode `\_12\catcode `\%12\relax}%
\providecommand \@@startlink[1]{}%
\providecommand \@@endlink[0]{}%
\providecommand \url  [0]{\begingroup\@sanitize@url \@url }%
\providecommand \@url [1]{\endgroup\@href {#1}{\urlprefix }}%
\providecommand \urlprefix  [0]{URL }%
\providecommand \Eprint [0]{\href }%
\providecommand \doibase [0]{https://doi.org/}%
\providecommand \selectlanguage [0]{\@gobble}%
\providecommand \bibinfo  [0]{\@secondoftwo}%
\providecommand \bibfield  [0]{\@secondoftwo}%
\providecommand \translation [1]{[#1]}%
\providecommand \BibitemOpen [0]{}%
\providecommand \bibitemStop [0]{}%
\providecommand \bibitemNoStop [0]{.\EOS\space}%
\providecommand \EOS [0]{\spacefactor3000\relax}%
\providecommand \BibitemShut  [1]{\csname bibitem#1\endcsname}%
\let\auto@bib@innerbib\@empty
%</preamble>
\bibitem [{\citenamefont {Damon}\ and\ \citenamefont
  {Eshbach}(1961)}]{damonMagnetostaticModesFerromagnet1961}%
  \BibitemOpen
  \bibfield  {author} {\bibinfo {author} {\bibfnamefont {R.~W.}\ \bibnamefont
  {Damon}}\ and\ \bibinfo {author} {\bibfnamefont {J.~R.}\ \bibnamefont
  {Eshbach}},\ }\href@noop {} {\bibfield  {journal} {\bibinfo  {journal} {J.
  Phys. Chem. Solids}\ }\textbf {\bibinfo {volume} {19}},\ \bibinfo {pages}
  {308} (\bibinfo {year} {1961})}\BibitemShut {NoStop}%
\bibitem [{\citenamefont
  {Kostylev}(2013)}]{kostylevNonreciprocityDipoleexchangeSpin2013}%
  \BibitemOpen
  \bibfield  {author} {\bibinfo {author} {\bibfnamefont {M.}~\bibnamefont
  {Kostylev}},\ }\href {https://doi.org/10.1063/1.4789962} {\bibfield
  {journal} {\bibinfo  {journal} {Journal of Applied Physics}\ }\textbf
  {\bibinfo {volume} {113}},\ \bibinfo {pages} {53907} (\bibinfo {year}
  {2013})}\BibitemShut {NoStop}%
\bibitem [{\citenamefont {Bogdanov}\ \emph {et~al.}(2002)\citenamefont
  {Bogdanov}, \citenamefont {R{\"o}{\ss}ler}, \citenamefont {Wolf},\ and\
  \citenamefont {M{\"u}ller}}]{bogdanovMagneticStructuresReorientation2002b}%
  \BibitemOpen
  \bibfield  {author} {\bibinfo {author} {\bibfnamefont {A.~N.}\ \bibnamefont
  {Bogdanov}}, \bibinfo {author} {\bibfnamefont {U.~K.}\ \bibnamefont
  {R{\"o}{\ss}ler}}, \bibinfo {author} {\bibfnamefont {M.}~\bibnamefont
  {Wolf}},\ and\ \bibinfo {author} {\bibfnamefont {K.-H.}\ \bibnamefont
  {M{\"u}ller}},\ }\href {https://doi.org/10.1103/PhysRevB.66.214410}
  {\bibfield  {journal} {\bibinfo  {journal} {Physical Review B}\ }\textbf
  {\bibinfo {volume} {66}},\ \bibinfo {pages} {214410} (\bibinfo {year}
  {2002})}\BibitemShut {NoStop}%
\bibitem [{\citenamefont
  {Dzyaloshinsky}(1958)}]{dzyaloshinsky_thermodynamic_1958}%
  \BibitemOpen
  \bibfield  {author} {\bibinfo {author} {\bibfnamefont {I.}~\bibnamefont
  {Dzyaloshinsky}},\ }\href {https://doi.org/10.1016/0022-3697(58)90076-3}
  {\bibfield  {journal} {\bibinfo  {journal} {Journal of Physics and Chemistry
  of Solids}\ }\textbf {\bibinfo {volume} {4}},\ \bibinfo {pages} {241}
  (\bibinfo {year} {1958})}\BibitemShut {NoStop}%
\bibitem [{\citenamefont {Moriya}(1960)}]{moriya_new_1960}%
  \BibitemOpen
  \bibfield  {author} {\bibinfo {author} {\bibfnamefont {T.}~\bibnamefont
  {Moriya}},\ }\href {https://doi.org/10.1103/PhysRevLett.4.228} {\bibfield
  {journal} {\bibinfo  {journal} {Physical Review Letters}\ }\textbf {\bibinfo
  {volume} {4}},\ \bibinfo {pages} {228} (\bibinfo {year} {1960})}\BibitemShut
  {NoStop}%
\bibitem [{\citenamefont {Nembach}\ \emph {et~al.}(2015)\citenamefont
  {Nembach}, \citenamefont {Shaw}, \citenamefont {Weiler}, \citenamefont
  {Ju{\'e}},\ and\ \citenamefont
  {Silva}}]{nembachLinearRelationHeisenberg2015}%
  \BibitemOpen
  \bibfield  {author} {\bibinfo {author} {\bibfnamefont {H.~T.}\ \bibnamefont
  {Nembach}}, \bibinfo {author} {\bibfnamefont {J.~M.}\ \bibnamefont {Shaw}},
  \bibinfo {author} {\bibfnamefont {M.}~\bibnamefont {Weiler}}, \bibinfo
  {author} {\bibfnamefont {E.}~\bibnamefont {Ju{\'e}}},\ and\ \bibinfo {author}
  {\bibfnamefont {T.~J.}\ \bibnamefont {Silva}},\ }\href
  {https://doi.org/10.1038/nphys3418} {\bibfield  {journal} {\bibinfo
  {journal} {Nat Phys}\ }\textbf {\bibinfo {volume} {11}},\ \bibinfo {pages}
  {825} (\bibinfo {year} {2015})}\BibitemShut {NoStop}%
\bibitem [{\citenamefont {Zakeri}\ \emph {et~al.}(2010)\citenamefont {Zakeri},
  \citenamefont {Zhang}, \citenamefont {Prokop}, \citenamefont {Chuang},
  \citenamefont {Sakr}, \citenamefont {Tang},\ and\ \citenamefont
  {Kirschner}}]{zakeriAsymmetricSpinWaveDispersion2010}%
  \BibitemOpen
  \bibfield  {author} {\bibinfo {author} {\bibfnamefont {K.}~\bibnamefont
  {Zakeri}}, \bibinfo {author} {\bibfnamefont {Y.}~\bibnamefont {Zhang}},
  \bibinfo {author} {\bibfnamefont {J.}~\bibnamefont {Prokop}}, \bibinfo
  {author} {\bibfnamefont {T.-H.}\ \bibnamefont {Chuang}}, \bibinfo {author}
  {\bibfnamefont {N.}~\bibnamefont {Sakr}}, \bibinfo {author} {\bibfnamefont
  {W.~X.}\ \bibnamefont {Tang}},\ and\ \bibinfo {author} {\bibfnamefont
  {J.}~\bibnamefont {Kirschner}},\ }\href
  {https://doi.org/10.1103/PhysRevLett.104.137203} {\bibfield  {journal}
  {\bibinfo  {journal} {Phys. Rev. Lett.}\ }\textbf {\bibinfo {volume} {104}},\
  \bibinfo {pages} {137203} (\bibinfo {year} {2010})}\BibitemShut {NoStop}%
\bibitem [{\citenamefont {{Cort{\'e}s-Ortu{\~n}o}}\ and\ \citenamefont
  {Landeros}(2013)}]{cortes-ortunoInfluenceDzyaloshinskiiMoriya2013}%
  \BibitemOpen
  \bibfield  {author} {\bibinfo {author} {\bibfnamefont {D.}~\bibnamefont
  {{Cort{\'e}s-Ortu{\~n}o}}}\ and\ \bibinfo {author} {\bibfnamefont
  {P.}~\bibnamefont {Landeros}},\ }\href
  {https://doi.org/10.1088/0953-8984/25/15/156001} {\bibfield  {journal}
  {\bibinfo  {journal} {Journal of Physics: Condensed Matter}\ }\textbf
  {\bibinfo {volume} {25}},\ \bibinfo {pages} {156001} (\bibinfo {year}
  {2013})}\BibitemShut {NoStop}%
\bibitem [{\citenamefont {Neusser}\ and\ \citenamefont
  {Grundler}(2009)}]{neusser_magnonics_2009}%
  \BibitemOpen
  \bibfield  {author} {\bibinfo {author} {\bibfnamefont {S.}~\bibnamefont
  {Neusser}}\ and\ \bibinfo {author} {\bibfnamefont {D.}~\bibnamefont
  {Grundler}},\ }\href {https://doi.org/10.1002/adma.200900809} {\bibfield
  {journal} {\bibinfo  {journal} {Advanced Materials}\ }\textbf {\bibinfo
  {volume} {21}},\ \bibinfo {pages} {2927} (\bibinfo {year}
  {2009})}\BibitemShut {NoStop}%
\bibitem [{\citenamefont {Chumak}\ \emph {et~al.}(2015)\citenamefont {Chumak},
  \citenamefont {Vasyuchka}, \citenamefont {Serga},\ and\ \citenamefont
  {Hillebrands}}]{chumak_magnon_2015}%
  \BibitemOpen
  \bibfield  {author} {\bibinfo {author} {\bibfnamefont {A.~V.}\ \bibnamefont
  {Chumak}}, \bibinfo {author} {\bibfnamefont {V.}~\bibnamefont {Vasyuchka}},
  \bibinfo {author} {\bibfnamefont {A.}~\bibnamefont {Serga}},\ and\ \bibinfo
  {author} {\bibfnamefont {B.}~\bibnamefont {Hillebrands}},\ }\href
  {https://doi.org/10.1038/nphys3347} {\bibfield  {journal} {\bibinfo
  {journal} {Nature Physics}\ }\textbf {\bibinfo {volume} {11}},\ \bibinfo
  {pages} {453} (\bibinfo {year} {2015})}\BibitemShut {NoStop}%
\bibitem [{\citenamefont {Chumak}\ \emph {et~al.}(2022)\citenamefont {Chumak},
  \citenamefont {Kabos}, \citenamefont {Wu}, \citenamefont {Abert},
  \citenamefont {Adelmann}, \citenamefont {Adeyeye}, \citenamefont {Akerman},
  \citenamefont {Aliev}, \citenamefont {Anane}, \citenamefont {Awad},
  \citenamefont {Back}, \citenamefont {Barman}, \citenamefont {Bauer},
  \citenamefont {Becherer}, \citenamefont {Beginin}, \citenamefont
  {Bittencourt}, \citenamefont {Blanter}, \citenamefont {Bortolotti},
  \citenamefont {Boventer}, \citenamefont {Bozhko}, \citenamefont {Bunyaev},
  \citenamefont {Carmiggelt}, \citenamefont {Cheenikundil}, \citenamefont
  {Ciubotaru}, \citenamefont {Cotofana}, \citenamefont {Csaba}, \citenamefont
  {Dobrovolskiy}, \citenamefont {Dubs}, \citenamefont {Elyasi}, \citenamefont
  {Fripp}, \citenamefont {Fulara}, \citenamefont {Golovchanskiy}, \citenamefont
  {Gonzalez-Ballestero}, \citenamefont {Graczyk}, \citenamefont {Grundler},
  \citenamefont {Gruszecki}, \citenamefont {Gubbiotti}, \citenamefont
  {Guslienko}, \citenamefont {Haldar}, \citenamefont {Hamdioui}, \citenamefont
  {Hertel}, \citenamefont {Hillebrands}, \citenamefont {Hioki}, \citenamefont
  {Houshang}, \citenamefont {Hu}, \citenamefont {Huebl}, \citenamefont {Huth},
  \citenamefont {Iacocca}, \citenamefont {Jungfleisch}, \citenamefont
  {Kakazei}, \citenamefont {Khitun}, \citenamefont {Khymyn}, \citenamefont
  {Kikkawa}, \citenamefont {Klaui}, \citenamefont {Klein}, \citenamefont
  {Klos}, \citenamefont {Knauer}, \citenamefont {Koraltan}, \citenamefont
  {Kostylev}, \citenamefont {Krawczyk}, \citenamefont {Krivorotov},
  \citenamefont {Kruglyak}, \citenamefont {Lachance-Quirion}, \citenamefont
  {Ladak}, \citenamefont {Lebrun}, \citenamefont {Li}, \citenamefont {Lindner},
  \citenamefont {Macedo}, \citenamefont {Mayr}, \citenamefont {Melkov},
  \citenamefont {Mieszczak}, \citenamefont {Nakamura}, \citenamefont {Nembach},
  \citenamefont {Nikitin}, \citenamefont {Nikitov}, \citenamefont {Novosad},
  \citenamefont {Otalora}, \citenamefont {Otani}, \citenamefont {Papp},
  \citenamefont {Pigeau}, \citenamefont {Pirro}, \citenamefont {Porod},
  \citenamefont {Porrati}, \citenamefont {Qin}, \citenamefont {Rana},
  \citenamefont {Reimann}, \citenamefont {Riente}, \citenamefont
  {Romero-Isart}, \citenamefont {Ross}, \citenamefont {Sadovnikov},
  \citenamefont {Safin}, \citenamefont {Saitoh}, \citenamefont {Schmidt},
  \citenamefont {Schultheiss}, \citenamefont {Schultheiss}, \citenamefont
  {Serga}, \citenamefont {Sharma}, \citenamefont {Shaw}, \citenamefont {Suess},
  \citenamefont {Surzhenko}, \citenamefont {Szulc}, \citenamefont {Taniguchi},
  \citenamefont {Urbanek}, \citenamefont {Usami}, \citenamefont {Ustinov},
  \citenamefont {van~der Sar}, \citenamefont {van Dijken}, \citenamefont
  {Vasyuchka}, \citenamefont {Verba}, \citenamefont {Kusminskiy}, \citenamefont
  {Wang}, \citenamefont {Weides}, \citenamefont {Weiler}, \citenamefont
  {Wintz}, \citenamefont {Wolski},\ and\ \citenamefont
  {Zhang}}]{chumak_advances_2022}%
  \BibitemOpen
  \bibfield  {author} {\bibinfo {author} {\bibfnamefont {A.~V.}\ \bibnamefont
  {Chumak}}, \bibinfo {author} {\bibfnamefont {P.}~\bibnamefont {Kabos}},
  \bibinfo {author} {\bibfnamefont {M.}~\bibnamefont {Wu}}, \bibinfo {author}
  {\bibfnamefont {C.}~\bibnamefont {Abert}}, \bibinfo {author} {\bibfnamefont
  {C.}~\bibnamefont {Adelmann}}, \bibinfo {author} {\bibfnamefont {A.~O.}\
  \bibnamefont {Adeyeye}}, \bibinfo {author} {\bibfnamefont {J.}~\bibnamefont
  {Akerman}}, \bibinfo {author} {\bibfnamefont {F.~G.}\ \bibnamefont {Aliev}},
  \bibinfo {author} {\bibfnamefont {A.}~\bibnamefont {Anane}}, \bibinfo
  {author} {\bibfnamefont {A.}~\bibnamefont {Awad}}, \bibinfo {author}
  {\bibfnamefont {C.~H.}\ \bibnamefont {Back}}, \bibinfo {author}
  {\bibfnamefont {A.}~\bibnamefont {Barman}}, \bibinfo {author} {\bibfnamefont
  {G.~E.~W.}\ \bibnamefont {Bauer}}, \bibinfo {author} {\bibfnamefont
  {M.}~\bibnamefont {Becherer}}, \bibinfo {author} {\bibfnamefont {E.~N.}\
  \bibnamefont {Beginin}}, \bibinfo {author} {\bibfnamefont {V.~A. S.~V.}\
  \bibnamefont {Bittencourt}}, \bibinfo {author} {\bibfnamefont {Y.~M.}\
  \bibnamefont {Blanter}}, \bibinfo {author} {\bibfnamefont {P.}~\bibnamefont
  {Bortolotti}}, \bibinfo {author} {\bibfnamefont {I.}~\bibnamefont
  {Boventer}}, \bibinfo {author} {\bibfnamefont {D.~A.}\ \bibnamefont
  {Bozhko}}, \bibinfo {author} {\bibfnamefont {S.~A.}\ \bibnamefont {Bunyaev}},
  \bibinfo {author} {\bibfnamefont {J.~J.}\ \bibnamefont {Carmiggelt}},
  \bibinfo {author} {\bibfnamefont {R.~R.}\ \bibnamefont {Cheenikundil}},
  \bibinfo {author} {\bibfnamefont {F.}~\bibnamefont {Ciubotaru}}, \bibinfo
  {author} {\bibfnamefont {S.}~\bibnamefont {Cotofana}}, \bibinfo {author}
  {\bibfnamefont {G.}~\bibnamefont {Csaba}}, \bibinfo {author} {\bibfnamefont
  {O.~V.}\ \bibnamefont {Dobrovolskiy}}, \bibinfo {author} {\bibfnamefont
  {C.}~\bibnamefont {Dubs}}, \bibinfo {author} {\bibfnamefont {M.}~\bibnamefont
  {Elyasi}}, \bibinfo {author} {\bibfnamefont {K.~G.}\ \bibnamefont {Fripp}},
  \bibinfo {author} {\bibfnamefont {H.}~\bibnamefont {Fulara}}, \bibinfo
  {author} {\bibfnamefont {I.~A.}\ \bibnamefont {Golovchanskiy}}, \bibinfo
  {author} {\bibfnamefont {C.}~\bibnamefont {Gonzalez-Ballestero}}, \bibinfo
  {author} {\bibfnamefont {P.}~\bibnamefont {Graczyk}}, \bibinfo {author}
  {\bibfnamefont {D.}~\bibnamefont {Grundler}}, \bibinfo {author}
  {\bibfnamefont {P.}~\bibnamefont {Gruszecki}}, \bibinfo {author}
  {\bibfnamefont {G.}~\bibnamefont {Gubbiotti}}, \bibinfo {author}
  {\bibfnamefont {K.}~\bibnamefont {Guslienko}}, \bibinfo {author}
  {\bibfnamefont {A.}~\bibnamefont {Haldar}}, \bibinfo {author} {\bibfnamefont
  {S.}~\bibnamefont {Hamdioui}}, \bibinfo {author} {\bibfnamefont
  {R.}~\bibnamefont {Hertel}}, \bibinfo {author} {\bibfnamefont
  {B.}~\bibnamefont {Hillebrands}}, \bibinfo {author} {\bibfnamefont
  {T.}~\bibnamefont {Hioki}}, \bibinfo {author} {\bibfnamefont
  {A.}~\bibnamefont {Houshang}}, \bibinfo {author} {\bibfnamefont {C.-M.}\
  \bibnamefont {Hu}}, \bibinfo {author} {\bibfnamefont {H.}~\bibnamefont
  {Huebl}}, \bibinfo {author} {\bibfnamefont {M.}~\bibnamefont {Huth}},
  \bibinfo {author} {\bibfnamefont {E.}~\bibnamefont {Iacocca}}, \bibinfo
  {author} {\bibfnamefont {M.~B.}\ \bibnamefont {Jungfleisch}}, \bibinfo
  {author} {\bibfnamefont {G.~N.}\ \bibnamefont {Kakazei}}, \bibinfo {author}
  {\bibfnamefont {A.}~\bibnamefont {Khitun}}, \bibinfo {author} {\bibfnamefont
  {R.}~\bibnamefont {Khymyn}}, \bibinfo {author} {\bibfnamefont
  {T.}~\bibnamefont {Kikkawa}}, \bibinfo {author} {\bibfnamefont
  {M.}~\bibnamefont {Klaui}}, \bibinfo {author} {\bibfnamefont
  {O.}~\bibnamefont {Klein}}, \bibinfo {author} {\bibfnamefont {J.~W.}\
  \bibnamefont {Klos}}, \bibinfo {author} {\bibfnamefont {S.}~\bibnamefont
  {Knauer}}, \bibinfo {author} {\bibfnamefont {S.}~\bibnamefont {Koraltan}},
  \bibinfo {author} {\bibfnamefont {M.}~\bibnamefont {Kostylev}}, \bibinfo
  {author} {\bibfnamefont {M.}~\bibnamefont {Krawczyk}}, \bibinfo {author}
  {\bibfnamefont {I.~N.}\ \bibnamefont {Krivorotov}}, \bibinfo {author}
  {\bibfnamefont {V.~V.}\ \bibnamefont {Kruglyak}}, \bibinfo {author}
  {\bibfnamefont {D.}~\bibnamefont {Lachance-Quirion}}, \bibinfo {author}
  {\bibfnamefont {S.}~\bibnamefont {Ladak}}, \bibinfo {author} {\bibfnamefont
  {R.}~\bibnamefont {Lebrun}}, \bibinfo {author} {\bibfnamefont
  {Y.}~\bibnamefont {Li}}, \bibinfo {author} {\bibfnamefont {M.}~\bibnamefont
  {Lindner}}, \bibinfo {author} {\bibfnamefont {R.}~\bibnamefont {Macedo}},
  \bibinfo {author} {\bibfnamefont {S.}~\bibnamefont {Mayr}}, \bibinfo {author}
  {\bibfnamefont {G.~A.}\ \bibnamefont {Melkov}}, \bibinfo {author}
  {\bibfnamefont {S.}~\bibnamefont {Mieszczak}}, \bibinfo {author}
  {\bibfnamefont {Y.}~\bibnamefont {Nakamura}}, \bibinfo {author}
  {\bibfnamefont {H.~T.}\ \bibnamefont {Nembach}}, \bibinfo {author}
  {\bibfnamefont {A.~A.}\ \bibnamefont {Nikitin}}, \bibinfo {author}
  {\bibfnamefont {S.~A.}\ \bibnamefont {Nikitov}}, \bibinfo {author}
  {\bibfnamefont {V.}~\bibnamefont {Novosad}}, \bibinfo {author} {\bibfnamefont
  {J.~A.}\ \bibnamefont {Otalora}}, \bibinfo {author} {\bibfnamefont
  {Y.}~\bibnamefont {Otani}}, \bibinfo {author} {\bibfnamefont
  {A.}~\bibnamefont {Papp}}, \bibinfo {author} {\bibfnamefont {B.}~\bibnamefont
  {Pigeau}}, \bibinfo {author} {\bibfnamefont {P.}~\bibnamefont {Pirro}},
  \bibinfo {author} {\bibfnamefont {W.}~\bibnamefont {Porod}}, \bibinfo
  {author} {\bibfnamefont {F.}~\bibnamefont {Porrati}}, \bibinfo {author}
  {\bibfnamefont {H.}~\bibnamefont {Qin}}, \bibinfo {author} {\bibfnamefont
  {B.}~\bibnamefont {Rana}}, \bibinfo {author} {\bibfnamefont {T.}~\bibnamefont
  {Reimann}}, \bibinfo {author} {\bibfnamefont {F.}~\bibnamefont {Riente}},
  \bibinfo {author} {\bibfnamefont {O.}~\bibnamefont {Romero-Isart}}, \bibinfo
  {author} {\bibfnamefont {A.}~\bibnamefont {Ross}}, \bibinfo {author}
  {\bibfnamefont {A.~V.}\ \bibnamefont {Sadovnikov}}, \bibinfo {author}
  {\bibfnamefont {A.~R.}\ \bibnamefont {Safin}}, \bibinfo {author}
  {\bibfnamefont {E.}~\bibnamefont {Saitoh}}, \bibinfo {author} {\bibfnamefont
  {G.}~\bibnamefont {Schmidt}}, \bibinfo {author} {\bibfnamefont
  {H.}~\bibnamefont {Schultheiss}}, \bibinfo {author} {\bibfnamefont
  {K.}~\bibnamefont {Schultheiss}}, \bibinfo {author} {\bibfnamefont {A.~A.}\
  \bibnamefont {Serga}}, \bibinfo {author} {\bibfnamefont {S.}~\bibnamefont
  {Sharma}}, \bibinfo {author} {\bibfnamefont {J.~M.}\ \bibnamefont {Shaw}},
  \bibinfo {author} {\bibfnamefont {D.}~\bibnamefont {Suess}}, \bibinfo
  {author} {\bibfnamefont {O.}~\bibnamefont {Surzhenko}}, \bibinfo {author}
  {\bibfnamefont {K.}~\bibnamefont {Szulc}}, \bibinfo {author} {\bibfnamefont
  {T.}~\bibnamefont {Taniguchi}}, \bibinfo {author} {\bibfnamefont
  {M.}~\bibnamefont {Urbanek}}, \bibinfo {author} {\bibfnamefont
  {K.}~\bibnamefont {Usami}}, \bibinfo {author} {\bibfnamefont {A.~B.}\
  \bibnamefont {Ustinov}}, \bibinfo {author} {\bibfnamefont {T.}~\bibnamefont
  {van~der Sar}}, \bibinfo {author} {\bibfnamefont {S.}~\bibnamefont {van
  Dijken}}, \bibinfo {author} {\bibfnamefont {V.~I.}\ \bibnamefont
  {Vasyuchka}}, \bibinfo {author} {\bibfnamefont {R.}~\bibnamefont {Verba}},
  \bibinfo {author} {\bibfnamefont {S.~V.}\ \bibnamefont {Kusminskiy}},
  \bibinfo {author} {\bibfnamefont {Q.}~\bibnamefont {Wang}}, \bibinfo {author}
  {\bibfnamefont {M.}~\bibnamefont {Weides}}, \bibinfo {author} {\bibfnamefont
  {M.}~\bibnamefont {Weiler}}, \bibinfo {author} {\bibfnamefont
  {S.}~\bibnamefont {Wintz}}, \bibinfo {author} {\bibfnamefont {S.~P.}\
  \bibnamefont {Wolski}},\ and\ \bibinfo {author} {\bibfnamefont
  {X.}~\bibnamefont {Zhang}},\ }\href
  {https://doi.org/10.1109/tmag.2022.3149664} {\bibfield  {journal} {\bibinfo
  {journal} {IEEE Transactions on Magnetics}\ }\textbf {\bibinfo {volume}
  {58}},\ \bibinfo {pages} {1–72} (\bibinfo {year} {2022})}\BibitemShut
  {NoStop}%
\bibitem [{\citenamefont {Flebus}\ \emph {et~al.}(2024)\citenamefont {Flebus},
  \citenamefont {Grundler}, \citenamefont {Rana}, \citenamefont {Otani},
  \citenamefont {Barsukov}, \citenamefont {Barman}, \citenamefont {Gubbiotti},
  \citenamefont {Landeros}, \citenamefont {Akerman}, \citenamefont {Ebels},
  \citenamefont {Pirro}, \citenamefont {Demidov}, \citenamefont {Schultheiss},
  \citenamefont {Csaba}, \citenamefont {Wang}, \citenamefont {Ciubotaru},
  \citenamefont {Nikonov}, \citenamefont {Che}, \citenamefont {Hertel},
  \citenamefont {Ono}, \citenamefont {Afanasiev}, \citenamefont {Mentink},
  \citenamefont {Rasing}, \citenamefont {Hillebrands}, \citenamefont
  {Kusminskiy}, \citenamefont {Zhang}, \citenamefont {Du}, \citenamefont
  {Finco}, \citenamefont {Van Der~Sar}, \citenamefont {Luo}, \citenamefont
  {Shiota}, \citenamefont {Sklenar}, \citenamefont {Yu},\ and\ \citenamefont
  {Rao}}]{flebus_2024_2024}%
  \BibitemOpen
  \bibfield  {author} {\bibinfo {author} {\bibfnamefont {B.}~\bibnamefont
  {Flebus}}, \bibinfo {author} {\bibfnamefont {D.}~\bibnamefont {Grundler}},
  \bibinfo {author} {\bibfnamefont {B.}~\bibnamefont {Rana}}, \bibinfo {author}
  {\bibfnamefont {Y.}~\bibnamefont {Otani}}, \bibinfo {author} {\bibfnamefont
  {I.}~\bibnamefont {Barsukov}}, \bibinfo {author} {\bibfnamefont
  {A.}~\bibnamefont {Barman}}, \bibinfo {author} {\bibfnamefont
  {G.}~\bibnamefont {Gubbiotti}}, \bibinfo {author} {\bibfnamefont
  {P.}~\bibnamefont {Landeros}}, \bibinfo {author} {\bibfnamefont
  {J.}~\bibnamefont {Akerman}}, \bibinfo {author} {\bibfnamefont
  {U.}~\bibnamefont {Ebels}}, \bibinfo {author} {\bibfnamefont
  {P.}~\bibnamefont {Pirro}}, \bibinfo {author} {\bibfnamefont {V.~E.}\
  \bibnamefont {Demidov}}, \bibinfo {author} {\bibfnamefont {K.}~\bibnamefont
  {Schultheiss}}, \bibinfo {author} {\bibfnamefont {G.}~\bibnamefont {Csaba}},
  \bibinfo {author} {\bibfnamefont {Q.}~\bibnamefont {Wang}}, \bibinfo {author}
  {\bibfnamefont {F.}~\bibnamefont {Ciubotaru}}, \bibinfo {author}
  {\bibfnamefont {D.~E.}\ \bibnamefont {Nikonov}}, \bibinfo {author}
  {\bibfnamefont {P.}~\bibnamefont {Che}}, \bibinfo {author} {\bibfnamefont
  {R.}~\bibnamefont {Hertel}}, \bibinfo {author} {\bibfnamefont
  {T.}~\bibnamefont {Ono}}, \bibinfo {author} {\bibfnamefont {D.}~\bibnamefont
  {Afanasiev}}, \bibinfo {author} {\bibfnamefont {J.}~\bibnamefont {Mentink}},
  \bibinfo {author} {\bibfnamefont {T.}~\bibnamefont {Rasing}}, \bibinfo
  {author} {\bibfnamefont {B.}~\bibnamefont {Hillebrands}}, \bibinfo {author}
  {\bibfnamefont {S.~V.}\ \bibnamefont {Kusminskiy}}, \bibinfo {author}
  {\bibfnamefont {W.}~\bibnamefont {Zhang}}, \bibinfo {author} {\bibfnamefont
  {C.~R.}\ \bibnamefont {Du}}, \bibinfo {author} {\bibfnamefont
  {A.}~\bibnamefont {Finco}}, \bibinfo {author} {\bibfnamefont
  {T.}~\bibnamefont {Van Der~Sar}}, \bibinfo {author} {\bibfnamefont {Y.~K.}\
  \bibnamefont {Luo}}, \bibinfo {author} {\bibfnamefont {Y.}~\bibnamefont
  {Shiota}}, \bibinfo {author} {\bibfnamefont {J.}~\bibnamefont {Sklenar}},
  \bibinfo {author} {\bibfnamefont {T.}~\bibnamefont {Yu}},\ and\ \bibinfo
  {author} {\bibfnamefont {J.}~\bibnamefont {Rao}},\ }\href
  {https://doi.org/10.1088/1361-648X/ad399c} {\bibfield  {journal} {\bibinfo
  {journal} {Journal of Physics: Condensed Matter}\ }\textbf {\bibinfo {volume}
  {36}},\ \bibinfo {pages} {363501} (\bibinfo {year} {2024})}\BibitemShut
  {NoStop}%
\bibitem [{\citenamefont {Hrabec}\ \emph {et~al.}(2017)\citenamefont {Hrabec},
  \citenamefont {Belmeguenai}, \citenamefont {Stashkevich}, \citenamefont
  {Ch{\'e}rif}, \citenamefont {Rohart}, \citenamefont {Roussign{\'e}},\ and\
  \citenamefont {Thiaville}}]{hrabecMakingDzyaloshinskiiMoriyaInteraction2017}%
  \BibitemOpen
  \bibfield  {author} {\bibinfo {author} {\bibfnamefont {A.}~\bibnamefont
  {Hrabec}}, \bibinfo {author} {\bibfnamefont {M.}~\bibnamefont {Belmeguenai}},
  \bibinfo {author} {\bibfnamefont {A.}~\bibnamefont {Stashkevich}}, \bibinfo
  {author} {\bibfnamefont {S.~M.}\ \bibnamefont {Ch{\'e}rif}}, \bibinfo
  {author} {\bibfnamefont {S.}~\bibnamefont {Rohart}}, \bibinfo {author}
  {\bibfnamefont {Y.}~\bibnamefont {Roussign{\'e}}},\ and\ \bibinfo {author}
  {\bibfnamefont {A.}~\bibnamefont {Thiaville}},\ }\href
  {https://doi.org/10.1063/1.4985649} {\bibfield  {journal} {\bibinfo
  {journal} {Applied Physics Letters}\ }\textbf {\bibinfo {volume} {110}},\
  \bibinfo {pages} {242402} (\bibinfo {year} {2017})}\BibitemShut {NoStop}%
\bibitem [{\citenamefont {Zhang}\ \emph {et~al.}(2022)\citenamefont {Zhang},
  \citenamefont {Liang}, \citenamefont {Bi}, \citenamefont {Zhao},
  \citenamefont {Bai}, \citenamefont {Cui}, \citenamefont {Zhou}, \citenamefont
  {Bai}, \citenamefont {Feng}, \citenamefont {Song}, \citenamefont {Chai},
  \citenamefont {Gladii}, \citenamefont {Schultheiss}, \citenamefont {Zhu},
  \citenamefont {Zhang}, \citenamefont {Peng}, \citenamefont {Yang},\ and\
  \citenamefont {Jiang}}]{zhang_quantifying_2022}%
  \BibitemOpen
  \bibfield  {author} {\bibinfo {author} {\bibfnamefont {Q.}~\bibnamefont
  {Zhang}}, \bibinfo {author} {\bibfnamefont {J.}~\bibnamefont {Liang}},
  \bibinfo {author} {\bibfnamefont {K.}~\bibnamefont {Bi}}, \bibinfo {author}
  {\bibfnamefont {L.}~\bibnamefont {Zhao}}, \bibinfo {author} {\bibfnamefont
  {H.}~\bibnamefont {Bai}}, \bibinfo {author} {\bibfnamefont {Q.}~\bibnamefont
  {Cui}}, \bibinfo {author} {\bibfnamefont {H.-A.}\ \bibnamefont {Zhou}},
  \bibinfo {author} {\bibfnamefont {H.}~\bibnamefont {Bai}}, \bibinfo {author}
  {\bibfnamefont {H.}~\bibnamefont {Feng}}, \bibinfo {author} {\bibfnamefont
  {W.}~\bibnamefont {Song}}, \bibinfo {author} {\bibfnamefont {G.}~\bibnamefont
  {Chai}}, \bibinfo {author} {\bibfnamefont {O.}~\bibnamefont {Gladii}},
  \bibinfo {author} {\bibfnamefont {H.}~\bibnamefont {Schultheiss}}, \bibinfo
  {author} {\bibfnamefont {T.}~\bibnamefont {Zhu}}, \bibinfo {author}
  {\bibfnamefont {J.}~\bibnamefont {Zhang}}, \bibinfo {author} {\bibfnamefont
  {Y.}~\bibnamefont {Peng}}, \bibinfo {author} {\bibfnamefont {H.}~\bibnamefont
  {Yang}},\ and\ \bibinfo {author} {\bibfnamefont {W.}~\bibnamefont {Jiang}},\
  }\href {https://doi.org/10.1103/PhysRevLett.128.167202} {\bibfield  {journal}
  {\bibinfo  {journal} {Physical Review Letters}\ }\textbf {\bibinfo {volume}
  {128}},\ \bibinfo {pages} {167202} (\bibinfo {year} {2022})}\BibitemShut
  {NoStop}%
\bibitem [{\citenamefont {Khalili~Amiri}\ \emph {et~al.}(2007)\citenamefont
  {Khalili~Amiri}, \citenamefont {Rejaei}, \citenamefont {Vroubel},\ and\
  \citenamefont {Zhuang}}]{khaliliamiriNonreciprocalSpinWave2007}%
  \BibitemOpen
  \bibfield  {author} {\bibinfo {author} {\bibfnamefont {P.}~\bibnamefont
  {Khalili~Amiri}}, \bibinfo {author} {\bibfnamefont {B.}~\bibnamefont
  {Rejaei}}, \bibinfo {author} {\bibfnamefont {M.}~\bibnamefont {Vroubel}},\
  and\ \bibinfo {author} {\bibfnamefont {Y.}~\bibnamefont {Zhuang}},\ }\href
  {https://doi.org/10.1063/1.2766842} {\bibfield  {journal} {\bibinfo
  {journal} {Applied Physics Letters}\ }\textbf {\bibinfo {volume} {91}},\
  \bibinfo {pages} {062502} (\bibinfo {year} {2007})}\BibitemShut {NoStop}%
\bibitem [{\citenamefont {Haidar}\ \emph {et~al.}(2014)\citenamefont {Haidar},
  \citenamefont {Bailleul}, \citenamefont {Kostylev},\ and\ \citenamefont
  {Lao}}]{haidarNonreciprocalOerstedField2014}%
  \BibitemOpen
  \bibfield  {author} {\bibinfo {author} {\bibfnamefont {M.}~\bibnamefont
  {Haidar}}, \bibinfo {author} {\bibfnamefont {M.}~\bibnamefont {Bailleul}},
  \bibinfo {author} {\bibfnamefont {M.}~\bibnamefont {Kostylev}},\ and\
  \bibinfo {author} {\bibfnamefont {Y.}~\bibnamefont {Lao}},\ }\href
  {https://doi.org/10.1103/PhysRevB.89.094426} {\bibfield  {journal} {\bibinfo
  {journal} {Physical Review B}\ }\textbf {\bibinfo {volume} {89}},\ \bibinfo
  {pages} {094426} (\bibinfo {year} {2014})}\BibitemShut {NoStop}%
\bibitem [{\citenamefont {Gladii}\ \emph {et~al.}(2016)\citenamefont {Gladii},
  \citenamefont {Haidar}, \citenamefont {Henry}, \citenamefont {Kostylev},\
  and\ \citenamefont {Bailleul}}]{gladiiFrequencyNonreciprocitySurface2016a}%
  \BibitemOpen
  \bibfield  {author} {\bibinfo {author} {\bibfnamefont {O.}~\bibnamefont
  {Gladii}}, \bibinfo {author} {\bibfnamefont {M.}~\bibnamefont {Haidar}},
  \bibinfo {author} {\bibfnamefont {Y.}~\bibnamefont {Henry}}, \bibinfo
  {author} {\bibfnamefont {M.}~\bibnamefont {Kostylev}},\ and\ \bibinfo
  {author} {\bibfnamefont {M.}~\bibnamefont {Bailleul}},\ }\href
  {https://doi.org/10.1103/PhysRevB.93.054430} {\bibfield  {journal} {\bibinfo
  {journal} {Physical Review B}\ }\textbf {\bibinfo {volume} {93}},\ \bibinfo
  {pages} {054430} (\bibinfo {year} {2016})}\BibitemShut {NoStop}%
\bibitem [{\citenamefont {Quessab}\ \emph {et~al.}(2020)\citenamefont
  {Quessab}, \citenamefont {Xu}, \citenamefont {Ma}, \citenamefont {Zhou},
  \citenamefont {Riley}, \citenamefont {Shaw}, \citenamefont {Nembach},
  \citenamefont {Poon},\ and\ \citenamefont
  {Kent}}]{quessabTuningInterfacialDzyaloshinskiiMoriya2020}%
  \BibitemOpen
  \bibfield  {author} {\bibinfo {author} {\bibfnamefont {Y.}~\bibnamefont
  {Quessab}}, \bibinfo {author} {\bibfnamefont {J.-W.}\ \bibnamefont {Xu}},
  \bibinfo {author} {\bibfnamefont {C.~T.}\ \bibnamefont {Ma}}, \bibinfo
  {author} {\bibfnamefont {W.}~\bibnamefont {Zhou}}, \bibinfo {author}
  {\bibfnamefont {G.~A.}\ \bibnamefont {Riley}}, \bibinfo {author}
  {\bibfnamefont {J.~M.}\ \bibnamefont {Shaw}}, \bibinfo {author}
  {\bibfnamefont {H.~T.}\ \bibnamefont {Nembach}}, \bibinfo {author}
  {\bibfnamefont {S.~J.}\ \bibnamefont {Poon}},\ and\ \bibinfo {author}
  {\bibfnamefont {A.~D.}\ \bibnamefont {Kent}},\ }\href
  {https://doi.org/10.1038/s41598-020-64427-0} {\bibfield  {journal} {\bibinfo
  {journal} {Scientific Reports}\ }\textbf {\bibinfo {volume} {10}},\ \bibinfo
  {pages} {7447} (\bibinfo {year} {2020})}\BibitemShut {NoStop}%
\bibitem [{\citenamefont {Ot{\'a}lora}\ \emph {et~al.}(2016)\citenamefont
  {Ot{\'a}lora}, \citenamefont {Yan}, \citenamefont {Schultheiss},
  \citenamefont {Hertel},\ and\ \citenamefont
  {K{\'a}kay}}]{otaloraCurvatureInducedAsymmetricSpinWave2016}%
  \BibitemOpen
  \bibfield  {author} {\bibinfo {author} {\bibfnamefont {J.~A.}\ \bibnamefont
  {Ot{\'a}lora}}, \bibinfo {author} {\bibfnamefont {M.}~\bibnamefont {Yan}},
  \bibinfo {author} {\bibfnamefont {H.}~\bibnamefont {Schultheiss}}, \bibinfo
  {author} {\bibfnamefont {R.}~\bibnamefont {Hertel}},\ and\ \bibinfo {author}
  {\bibfnamefont {A.}~\bibnamefont {K{\'a}kay}},\ }\href
  {https://doi.org/10.1103/PhysRevLett.117.227203} {\bibfield  {journal}
  {\bibinfo  {journal} {Physical Review Letters}\ }\textbf {\bibinfo {volume}
  {117}},\ \bibinfo {pages} {227203} (\bibinfo {year} {2016})}\BibitemShut
  {NoStop}%
\bibitem [{\citenamefont {Ot{\'a}lora}\ \emph {et~al.}(2017)\citenamefont
  {Ot{\'a}lora}, \citenamefont {Yan}, \citenamefont {Schultheiss},
  \citenamefont {Hertel},\ and\ \citenamefont
  {K{\'a}kay}}]{otaloraAsymmetricSpinwaveDispersion2017}%
  \BibitemOpen
  \bibfield  {author} {\bibinfo {author} {\bibfnamefont {J.~A.}\ \bibnamefont
  {Ot{\'a}lora}}, \bibinfo {author} {\bibfnamefont {M.}~\bibnamefont {Yan}},
  \bibinfo {author} {\bibfnamefont {H.}~\bibnamefont {Schultheiss}}, \bibinfo
  {author} {\bibfnamefont {R.}~\bibnamefont {Hertel}},\ and\ \bibinfo {author}
  {\bibfnamefont {A.}~\bibnamefont {K{\'a}kay}},\ }\href
  {https://doi.org/10.1103/PhysRevB.95.184415} {\bibfield  {journal} {\bibinfo
  {journal} {PHYSICAL REVIEW B}\ }\textbf {\bibinfo {volume} {95}},\ \bibinfo
  {pages} {184415} (\bibinfo {year} {2017})}\BibitemShut {NoStop}%
\bibitem [{\citenamefont {{Salazar-Cardona}}\ \emph {et~al.}(2021)\citenamefont
  {{Salazar-Cardona}}, \citenamefont {K{\"o}rber}, \citenamefont {Schultheiss},
  \citenamefont {Lenz}, \citenamefont {Thomas}, \citenamefont {Nielsch},
  \citenamefont {K{\'a}kay},\ and\ \citenamefont
  {Ot{\'a}lora}}]{salazar-cardonaNonreciprocitySpinWaves2021}%
  \BibitemOpen
  \bibfield  {author} {\bibinfo {author} {\bibfnamefont {M.~M.}\ \bibnamefont
  {{Salazar-Cardona}}}, \bibinfo {author} {\bibfnamefont {L.}~\bibnamefont
  {K{\"o}rber}}, \bibinfo {author} {\bibfnamefont {H.}~\bibnamefont
  {Schultheiss}}, \bibinfo {author} {\bibfnamefont {K.}~\bibnamefont {Lenz}},
  \bibinfo {author} {\bibfnamefont {A.}~\bibnamefont {Thomas}}, \bibinfo
  {author} {\bibfnamefont {K.}~\bibnamefont {Nielsch}}, \bibinfo {author}
  {\bibfnamefont {A.}~\bibnamefont {K{\'a}kay}},\ and\ \bibinfo {author}
  {\bibfnamefont {J.~A.}\ \bibnamefont {Ot{\'a}lora}},\ }\href
  {https://doi.org/10.1063/5.0048692} {\bibfield  {journal} {\bibinfo
  {journal} {Applied Physics Letters}\ }\textbf {\bibinfo {volume} {118}},\
  \bibinfo {pages} {262411} (\bibinfo {year} {2021})}\BibitemShut {NoStop}%
\bibitem [{\citenamefont {Cho}\ \emph {et~al.}(2015)\citenamefont {Cho},
  \citenamefont {Kim}, \citenamefont {Lee}, \citenamefont {Kim}, \citenamefont
  {Lavrijsen}, \citenamefont {Solignac}, \citenamefont {Yin}, \citenamefont
  {Han}, \citenamefont {van Hoof}, \citenamefont {Swagten}, \citenamefont
  {Koopmans},\ and\ \citenamefont {You}}]{cho_thickness_2015}%
  \BibitemOpen
  \bibfield  {author} {\bibinfo {author} {\bibfnamefont {J.}~\bibnamefont
  {Cho}}, \bibinfo {author} {\bibfnamefont {N.-H.}\ \bibnamefont {Kim}},
  \bibinfo {author} {\bibfnamefont {S.}~\bibnamefont {Lee}}, \bibinfo {author}
  {\bibfnamefont {J.-S.}\ \bibnamefont {Kim}}, \bibinfo {author} {\bibfnamefont
  {R.}~\bibnamefont {Lavrijsen}}, \bibinfo {author} {\bibfnamefont
  {A.}~\bibnamefont {Solignac}}, \bibinfo {author} {\bibfnamefont
  {Y.}~\bibnamefont {Yin}}, \bibinfo {author} {\bibfnamefont {D.-S.}\
  \bibnamefont {Han}}, \bibinfo {author} {\bibfnamefont {N.~J.~J.}\
  \bibnamefont {van Hoof}}, \bibinfo {author} {\bibfnamefont {H.~J.~M.}\
  \bibnamefont {Swagten}}, \bibinfo {author} {\bibfnamefont {B.}~\bibnamefont
  {Koopmans}},\ and\ \bibinfo {author} {\bibfnamefont {C.-Y.}\ \bibnamefont
  {You}},\ }\href {https://doi.org/10.1038/ncomms8635} {\bibfield  {journal}
  {\bibinfo  {journal} {Nature Communications}\ }\textbf {\bibinfo {volume}
  {6}},\ \bibinfo {pages} {7635} (\bibinfo {year} {2015})}\BibitemShut
  {NoStop}%
\bibitem [{\citenamefont {Gallardo}\ \emph
  {et~al.}(2019{\natexlab{a}})\citenamefont {Gallardo}, \citenamefont
  {Schneider}, \citenamefont {Chaurasiya}, \citenamefont {Oelschl{\"a}gel},
  \citenamefont {Arekapudi}, \citenamefont {{Rold{\'a}n-Molina}}, \citenamefont
  {H{\"u}bner}, \citenamefont {Lenz}, \citenamefont {Barman}, \citenamefont
  {Fassbender}, \citenamefont {Lindner}, \citenamefont {Hellwig},\ and\
  \citenamefont {Landeros}}]{gallardoReconfigurableSpinWaveNonreciprocity2019}%
  \BibitemOpen
  \bibfield  {author} {\bibinfo {author} {\bibfnamefont {R.~A.}\ \bibnamefont
  {Gallardo}}, \bibinfo {author} {\bibfnamefont {T.}~\bibnamefont {Schneider}},
  \bibinfo {author} {\bibfnamefont {A.~K.}\ \bibnamefont {Chaurasiya}},
  \bibinfo {author} {\bibfnamefont {A.}~\bibnamefont {Oelschl{\"a}gel}},
  \bibinfo {author} {\bibfnamefont {S.~S.}\ \bibnamefont {Arekapudi}}, \bibinfo
  {author} {\bibfnamefont {A.}~\bibnamefont {{Rold{\'a}n-Molina}}}, \bibinfo
  {author} {\bibfnamefont {R.}~\bibnamefont {H{\"u}bner}}, \bibinfo {author}
  {\bibfnamefont {K.}~\bibnamefont {Lenz}}, \bibinfo {author} {\bibfnamefont
  {A.}~\bibnamefont {Barman}}, \bibinfo {author} {\bibfnamefont
  {J.}~\bibnamefont {Fassbender}}, \bibinfo {author} {\bibfnamefont
  {J.}~\bibnamefont {Lindner}}, \bibinfo {author} {\bibfnamefont
  {O.}~\bibnamefont {Hellwig}},\ and\ \bibinfo {author} {\bibfnamefont
  {P.}~\bibnamefont {Landeros}},\ }\href
  {https://doi.org/10.1103/PhysRevApplied.12.034012} {\bibfield  {journal}
  {\bibinfo  {journal} {Physical Review Applied}\ }\textbf {\bibinfo {volume}
  {12}},\ \bibinfo {pages} {034012} (\bibinfo {year}
  {2019}{\natexlab{a}})}\BibitemShut {NoStop}%
\bibitem [{\citenamefont {Gallardo}\ \emph
  {et~al.}(2019{\natexlab{b}})\citenamefont {Gallardo}, \citenamefont
  {{Alvarado-Seguel}}, \citenamefont {Schneider}, \citenamefont
  {{Gonzalez-Fuentes}}, \citenamefont {{Rold{\'a}n-Molina}}, \citenamefont
  {Lenz}, \citenamefont {Lindner},\ and\ \citenamefont
  {Landeros}}]{gallardoSpinwaveNonreciprocityMagnetizationgraded2019}%
  \BibitemOpen
  \bibfield  {author} {\bibinfo {author} {\bibfnamefont {R.~A.}\ \bibnamefont
  {Gallardo}}, \bibinfo {author} {\bibfnamefont {P.}~\bibnamefont
  {{Alvarado-Seguel}}}, \bibinfo {author} {\bibfnamefont {T.}~\bibnamefont
  {Schneider}}, \bibinfo {author} {\bibfnamefont {C.}~\bibnamefont
  {{Gonzalez-Fuentes}}}, \bibinfo {author} {\bibfnamefont {A.}~\bibnamefont
  {{Rold{\'a}n-Molina}}}, \bibinfo {author} {\bibfnamefont {K.}~\bibnamefont
  {Lenz}}, \bibinfo {author} {\bibfnamefont {J.}~\bibnamefont {Lindner}},\ and\
  \bibinfo {author} {\bibfnamefont {P.}~\bibnamefont {Landeros}},\ }\href
  {https://doi.org/10.1088/1367-2630/ab0449} {\bibfield  {journal} {\bibinfo
  {journal} {New Journal of Physics}\ }\textbf {\bibinfo {volume} {21}},\
  \bibinfo {pages} {033026} (\bibinfo {year} {2019}{\natexlab{b}})}\BibitemShut
  {NoStop}%
\bibitem [{\citenamefont {Henry}\ \emph {et~al.}(2016)\citenamefont {Henry},
  \citenamefont {Gladii},\ and\ \citenamefont
  {Bailleul}}]{henryPropagatingSpinwaveNormal2016}%
  \BibitemOpen
  \bibfield  {author} {\bibinfo {author} {\bibfnamefont {Y.}~\bibnamefont
  {Henry}}, \bibinfo {author} {\bibfnamefont {O.}~\bibnamefont {Gladii}},\ and\
  \bibinfo {author} {\bibfnamefont {M.}~\bibnamefont {Bailleul}},\ }\href@noop
  {} {\bibfield  {journal} {\bibinfo  {journal} {arXiv:1611.06153 [cond-mat]}\
  } (\bibinfo {year} {2016})},\ \Eprint {https://arxiv.org/abs/1611.06153}
  {arXiv:1611.06153 [cond-mat]} \BibitemShut {NoStop}%
\bibitem [{\citenamefont {Grassi}\ \emph {et~al.}(2020)\citenamefont {Grassi},
  \citenamefont {Geilen}, \citenamefont {Louis}, \citenamefont {Mohseni},
  \citenamefont {Br{\"a}cher}, \citenamefont {Hehn}, \citenamefont {Stoeffler},
  \citenamefont {Bailleul}, \citenamefont {Pirro},\ and\ \citenamefont
  {Henry}}]{grassiSlowWaveBasedNanomagnonicDiode2020}%
  \BibitemOpen
  \bibfield  {author} {\bibinfo {author} {\bibfnamefont {M.}~\bibnamefont
  {Grassi}}, \bibinfo {author} {\bibfnamefont {M.}~\bibnamefont {Geilen}},
  \bibinfo {author} {\bibfnamefont {D.}~\bibnamefont {Louis}}, \bibinfo
  {author} {\bibfnamefont {M.}~\bibnamefont {Mohseni}}, \bibinfo {author}
  {\bibfnamefont {T.}~\bibnamefont {Br{\"a}cher}}, \bibinfo {author}
  {\bibfnamefont {M.}~\bibnamefont {Hehn}}, \bibinfo {author} {\bibfnamefont
  {D.}~\bibnamefont {Stoeffler}}, \bibinfo {author} {\bibfnamefont
  {M.}~\bibnamefont {Bailleul}}, \bibinfo {author} {\bibfnamefont
  {P.}~\bibnamefont {Pirro}},\ and\ \bibinfo {author} {\bibfnamefont
  {Y.}~\bibnamefont {Henry}},\ }\href
  {https://doi.org/10.1103/PhysRevApplied.14.024047} {\bibfield  {journal}
  {\bibinfo  {journal} {Physical Review Applied}\ }\textbf {\bibinfo {volume}
  {14}},\ \bibinfo {pages} {024047} (\bibinfo {year} {2020})}\BibitemShut
  {NoStop}%
\bibitem [{\citenamefont {Sluka}\ \emph {et~al.}(2019)\citenamefont {Sluka},
  \citenamefont {Schneider}, \citenamefont {Gallardo}, \citenamefont
  {K{\'a}kay}, \citenamefont {Weigand}, \citenamefont {Warnatz}, \citenamefont
  {Mattheis}, \citenamefont {{Rold{\'a}n-Molina}}, \citenamefont {Landeros},
  \citenamefont {Tiberkevich}, \citenamefont {Slavin}, \citenamefont
  {Sch{\"u}tz}, \citenamefont {Erbe}, \citenamefont {Deac}, \citenamefont
  {Lindner}, \citenamefont {Raabe}, \citenamefont {Fassbender},\ and\
  \citenamefont {Wintz}}]{slukaEmissionPropagation1D2019}%
  \BibitemOpen
  \bibfield  {author} {\bibinfo {author} {\bibfnamefont {V.}~\bibnamefont
  {Sluka}}, \bibinfo {author} {\bibfnamefont {T.}~\bibnamefont {Schneider}},
  \bibinfo {author} {\bibfnamefont {R.~A.}\ \bibnamefont {Gallardo}}, \bibinfo
  {author} {\bibfnamefont {A.}~\bibnamefont {K{\'a}kay}}, \bibinfo {author}
  {\bibfnamefont {M.}~\bibnamefont {Weigand}}, \bibinfo {author} {\bibfnamefont
  {T.}~\bibnamefont {Warnatz}}, \bibinfo {author} {\bibfnamefont
  {R.}~\bibnamefont {Mattheis}}, \bibinfo {author} {\bibfnamefont
  {A.}~\bibnamefont {{Rold{\'a}n-Molina}}}, \bibinfo {author} {\bibfnamefont
  {P.}~\bibnamefont {Landeros}}, \bibinfo {author} {\bibfnamefont
  {V.}~\bibnamefont {Tiberkevich}}, \bibinfo {author} {\bibfnamefont
  {A.}~\bibnamefont {Slavin}}, \bibinfo {author} {\bibfnamefont
  {G.}~\bibnamefont {Sch{\"u}tz}}, \bibinfo {author} {\bibfnamefont
  {A.}~\bibnamefont {Erbe}}, \bibinfo {author} {\bibfnamefont {A.}~\bibnamefont
  {Deac}}, \bibinfo {author} {\bibfnamefont {J.}~\bibnamefont {Lindner}},
  \bibinfo {author} {\bibfnamefont {J.}~\bibnamefont {Raabe}}, \bibinfo
  {author} {\bibfnamefont {J.}~\bibnamefont {Fassbender}},\ and\ \bibinfo
  {author} {\bibfnamefont {S.}~\bibnamefont {Wintz}},\ }\href
  {https://doi.org/10.1038/s41565-019-0383-4} {\bibfield  {journal} {\bibinfo
  {journal} {Nature Nanotechnology}\ }\textbf {\bibinfo {volume} {14}},\
  \bibinfo {pages} {328} (\bibinfo {year} {2019})}\BibitemShut {NoStop}%
\bibitem [{\citenamefont {Shichi}\ \emph {et~al.}(2015)\citenamefont {Shichi},
  \citenamefont {Kanazawa}, \citenamefont {Matsuda}, \citenamefont {Okajima},
  \citenamefont {Hasegawa}, \citenamefont {Okada}, \citenamefont {Goto},
  \citenamefont {Takagi},\ and\ \citenamefont
  {Inoue}}]{shichiSpinWaveIsolator2015}%
  \BibitemOpen
  \bibfield  {author} {\bibinfo {author} {\bibfnamefont {S.}~\bibnamefont
  {Shichi}}, \bibinfo {author} {\bibfnamefont {N.}~\bibnamefont {Kanazawa}},
  \bibinfo {author} {\bibfnamefont {K.}~\bibnamefont {Matsuda}}, \bibinfo
  {author} {\bibfnamefont {S.}~\bibnamefont {Okajima}}, \bibinfo {author}
  {\bibfnamefont {T.}~\bibnamefont {Hasegawa}}, \bibinfo {author}
  {\bibfnamefont {T.}~\bibnamefont {Okada}}, \bibinfo {author} {\bibfnamefont
  {T.}~\bibnamefont {Goto}}, \bibinfo {author} {\bibfnamefont {H.}~\bibnamefont
  {Takagi}},\ and\ \bibinfo {author} {\bibfnamefont {M.}~\bibnamefont
  {Inoue}},\ }\href {https://doi.org/10.1063/1.4915101} {\bibfield  {journal}
  {\bibinfo  {journal} {Journal of Applied Physics}\ }\textbf {\bibinfo
  {volume} {117}},\ \bibinfo {pages} {17D125} (\bibinfo {year}
  {2015})}\BibitemShut {NoStop}%
\bibitem [{\citenamefont {Thiancourt}\ \emph {et~al.}(2024)\citenamefont
  {Thiancourt}, \citenamefont {Ngom}, \citenamefont {Bardou},\ and\
  \citenamefont {Devolder}}]{thiancourtUnidirectionalSpinWaves2024}%
  \BibitemOpen
  \bibfield  {author} {\bibinfo {author} {\bibfnamefont {G.}~\bibnamefont
  {Thiancourt}}, \bibinfo {author} {\bibfnamefont {S.}~\bibnamefont {Ngom}},
  \bibinfo {author} {\bibfnamefont {N.}~\bibnamefont {Bardou}},\ and\ \bibinfo
  {author} {\bibfnamefont {T.}~\bibnamefont {Devolder}},\ }\href
  {https://doi.org/10.1103/PhysRevApplied.22.034040} {\bibfield  {journal}
  {\bibinfo  {journal} {Physical Review Applied}\ }\textbf {\bibinfo {volume}
  {22}},\ \bibinfo {pages} {034040} (\bibinfo {year} {2024})}\BibitemShut
  {NoStop}%
\bibitem [{\citenamefont {Wojewoda}\ \emph {et~al.}(2024)\citenamefont
  {Wojewoda}, \citenamefont {Holobr{\'a}dek}, \citenamefont {Pavelka},
  \citenamefont {Pribytova}, \citenamefont {Kr{\v c}ma}, \citenamefont
  {Kl{\'i}ma}, \citenamefont {Panda}, \citenamefont {Michali{\v c}ka},
  \citenamefont {Lednick{\'y}}, \citenamefont {Chumak},\ and\ \citenamefont
  {Urb{\'a}nek}}]{wojewodaUnidirectionalPropagationZeromomentum2024}%
  \BibitemOpen
  \bibfield  {author} {\bibinfo {author} {\bibfnamefont {O.}~\bibnamefont
  {Wojewoda}}, \bibinfo {author} {\bibfnamefont {J.}~\bibnamefont
  {Holobr{\'a}dek}}, \bibinfo {author} {\bibfnamefont {D.}~\bibnamefont
  {Pavelka}}, \bibinfo {author} {\bibfnamefont {E.}~\bibnamefont {Pribytova}},
  \bibinfo {author} {\bibfnamefont {J.}~\bibnamefont {Kr{\v c}ma}}, \bibinfo
  {author} {\bibfnamefont {J.}~\bibnamefont {Kl{\'i}ma}}, \bibinfo {author}
  {\bibfnamefont {J.}~\bibnamefont {Panda}}, \bibinfo {author} {\bibfnamefont
  {J.}~\bibnamefont {Michali{\v c}ka}}, \bibinfo {author} {\bibfnamefont
  {T.}~\bibnamefont {Lednick{\'y}}}, \bibinfo {author} {\bibfnamefont {A.~V.}\
  \bibnamefont {Chumak}},\ and\ \bibinfo {author} {\bibfnamefont
  {M.}~\bibnamefont {Urb{\'a}nek}},\ }\href {https://doi.org/10.1063/5.0218478}
  {\bibfield  {journal} {\bibinfo  {journal} {Applied Physics Letters}\
  }\textbf {\bibinfo {volume} {125}},\ \bibinfo {pages} {132401} (\bibinfo
  {year} {2024})}\BibitemShut {NoStop}%
\bibitem [{\citenamefont {K\"{o}rber}\ \emph {et~al.}(2022)\citenamefont
  {K\"{o}rber}, \citenamefont {Hempel}, \citenamefont {Otto}, \citenamefont
  {Gallardo}, \citenamefont {Henry}, \citenamefont {Lindner},\ and\
  \citenamefont {Kákay}}]{korberFiniteelementDynamicmatrixApproach2022}%
  \BibitemOpen
  \bibfield  {author} {\bibinfo {author} {\bibfnamefont {L.}~\bibnamefont
  {K\"{o}rber}}, \bibinfo {author} {\bibfnamefont {A.}~\bibnamefont {Hempel}},
  \bibinfo {author} {\bibfnamefont {A.}~\bibnamefont {Otto}}, \bibinfo {author}
  {\bibfnamefont {R.~A.}\ \bibnamefont {Gallardo}}, \bibinfo {author}
  {\bibfnamefont {Y.}~\bibnamefont {Henry}}, \bibinfo {author} {\bibfnamefont
  {J.}~\bibnamefont {Lindner}},\ and\ \bibinfo {author} {\bibfnamefont
  {A.}~\bibnamefont {Kákay}},\ }\bibfield  {journal} {\bibinfo  {journal} {AIP
  Advances}\ }\textbf {\bibinfo {volume} {12}},\ \href
  {https://doi.org/10.1063/5.0107457} {10.1063/5.0107457} (\bibinfo {year}
  {2022})\BibitemShut {NoStop}%
\bibitem [{\citenamefont {K{\"o}rber}\ \emph {et~al.}(2022)\citenamefont
  {K{\"o}rber}, \citenamefont {Quasebarth}, \citenamefont {Hempel},
  \citenamefont {Zahn}, \citenamefont {Andreas}, \citenamefont {Westphal},
  \citenamefont {Hertel},\ and\ \citenamefont
  {K{\'a}kay}}]{korberTetraXFiniteElementMicromagneticModeling2022}%
  \BibitemOpen
  \bibfield  {author} {\bibinfo {author} {\bibfnamefont {L.}~\bibnamefont
  {K{\"o}rber}}, \bibinfo {author} {\bibfnamefont {G.}~\bibnamefont
  {Quasebarth}}, \bibinfo {author} {\bibfnamefont {A.}~\bibnamefont {Hempel}},
  \bibinfo {author} {\bibfnamefont {F.}~\bibnamefont {Zahn}}, \bibinfo {author}
  {\bibfnamefont {O.}~\bibnamefont {Andreas}}, \bibinfo {author} {\bibfnamefont
  {E.}~\bibnamefont {Westphal}}, \bibinfo {author} {\bibfnamefont
  {R.}~\bibnamefont {Hertel}},\ and\ \bibinfo {author} {\bibfnamefont
  {A.}~\bibnamefont {K{\'a}kay}},\ }\href
  {https://doi.org/10.14278/rodare.1418} {\bibinfo {title} {{{TetraX}}:
  {{Finite-Element Micromagnetic-Modeling Package}}}} (\bibinfo {year}
  {2022})\BibitemShut {NoStop}%
\bibitem [{\citenamefont {Tacchi}\ \emph {et~al.}(2019)\citenamefont {Tacchi},
  \citenamefont {Silvani}, \citenamefont {Carlotti}, \citenamefont {Marangolo},
  \citenamefont {Eddrief}, \citenamefont {Rettori},\ and\ \citenamefont
  {Pini}}]{tacchiStronglyHybridizedDipoleexchange2019}%
  \BibitemOpen
  \bibfield  {author} {\bibinfo {author} {\bibfnamefont {S.}~\bibnamefont
  {Tacchi}}, \bibinfo {author} {\bibfnamefont {R.}~\bibnamefont {Silvani}},
  \bibinfo {author} {\bibfnamefont {G.}~\bibnamefont {Carlotti}}, \bibinfo
  {author} {\bibfnamefont {M.}~\bibnamefont {Marangolo}}, \bibinfo {author}
  {\bibfnamefont {M.}~\bibnamefont {Eddrief}}, \bibinfo {author} {\bibfnamefont
  {A.}~\bibnamefont {Rettori}},\ and\ \bibinfo {author} {\bibfnamefont {M.~G.}\
  \bibnamefont {Pini}},\ }\href {https://doi.org/10.1103/PhysRevB.100.104406}
  {\bibfield  {journal} {\bibinfo  {journal} {Physical Review B}\ }\textbf
  {\bibinfo {volume} {100}},\ \bibinfo {pages} {104406} (\bibinfo {year}
  {2019})}\BibitemShut {NoStop}%
\bibitem [{\citenamefont {Klingler}\ \emph {et~al.}(2018)\citenamefont
  {Klingler}, \citenamefont {Amin}, \citenamefont {Gepr{\"a}gs}, \citenamefont
  {Ganzhorn}, \citenamefont {Maier-Flaig}, \citenamefont {Althammer},
  \citenamefont {Huebl}, \citenamefont {Gross}, \citenamefont {McMichael},
  \citenamefont {Stiles}, \citenamefont {Goennenwein},\ and\ \citenamefont
  {Weiler}}]{Klingler2018-np}%
  \BibitemOpen
  \bibfield  {author} {\bibinfo {author} {\bibfnamefont {S.}~\bibnamefont
  {Klingler}}, \bibinfo {author} {\bibfnamefont {V.}~\bibnamefont {Amin}},
  \bibinfo {author} {\bibfnamefont {S.}~\bibnamefont {Gepr{\"a}gs}}, \bibinfo
  {author} {\bibfnamefont {K.}~\bibnamefont {Ganzhorn}}, \bibinfo {author}
  {\bibfnamefont {H.}~\bibnamefont {Maier-Flaig}}, \bibinfo {author}
  {\bibfnamefont {M.}~\bibnamefont {Althammer}}, \bibinfo {author}
  {\bibfnamefont {H.}~\bibnamefont {Huebl}}, \bibinfo {author} {\bibfnamefont
  {R.}~\bibnamefont {Gross}}, \bibinfo {author} {\bibfnamefont {R.~D.}\
  \bibnamefont {McMichael}}, \bibinfo {author} {\bibfnamefont {M.~D.}\
  \bibnamefont {Stiles}}, \bibinfo {author} {\bibfnamefont {S.~T.~B.}\
  \bibnamefont {Goennenwein}},\ and\ \bibinfo {author} {\bibfnamefont
  {M.}~\bibnamefont {Weiler}},\ }\href@noop {} {\bibfield  {journal} {\bibinfo
  {journal} {Phys. Rev. Lett.}\ }\textbf {\bibinfo {volume} {120}},\ \bibinfo
  {pages} {127201} (\bibinfo {year} {2018})}\BibitemShut {NoStop}%
\bibitem [{\citenamefont {K{\"o}rber}\ \emph {et~al.}(2021)\citenamefont
  {K{\"o}rber}, \citenamefont {Quasebarth}, \citenamefont {Otto},\ and\
  \citenamefont {K{\'a}kay}}]{korberFiniteelementDynamicmatrixApproach2021}%
  \BibitemOpen
  \bibfield  {author} {\bibinfo {author} {\bibfnamefont {L.}~\bibnamefont
  {K{\"o}rber}}, \bibinfo {author} {\bibfnamefont {G.}~\bibnamefont
  {Quasebarth}}, \bibinfo {author} {\bibfnamefont {A.}~\bibnamefont {Otto}},\
  and\ \bibinfo {author} {\bibfnamefont {A.}~\bibnamefont {K{\'a}kay}},\ }\href
  {https://doi.org/10.1063/5.0054169} {\bibfield  {journal} {\bibinfo
  {journal} {AIP Advances}\ }\textbf {\bibinfo {volume} {11}},\ \bibinfo
  {pages} {095006} (\bibinfo {year} {2021})}\BibitemShut {NoStop}%
\bibitem [{\citenamefont {Demokritov}\ \emph {et~al.}(2001)\citenamefont
  {Demokritov}, \citenamefont {Hillebrands},\ and\ \citenamefont
  {Slavin}}]{demokritovBrillouinLightScattering2001}%
  \BibitemOpen
  \bibfield  {author} {\bibinfo {author} {\bibfnamefont {S.~O.}\ \bibnamefont
  {Demokritov}}, \bibinfo {author} {\bibfnamefont {B.}~\bibnamefont
  {Hillebrands}},\ and\ \bibinfo {author} {\bibfnamefont {A.~N.}\ \bibnamefont
  {Slavin}},\ }\href@noop {} {\bibfield  {journal} {\bibinfo  {journal}
  {Physics Reports}\ }\textbf {\bibinfo {volume} {348}},\ \bibinfo {pages}
  {441} (\bibinfo {year} {2001})}\BibitemShut {NoStop}%
\bibitem [{\citenamefont {Sebastian}\ \emph {et~al.}(2015)\citenamefont
  {Sebastian}, \citenamefont {Schultheiss}, \citenamefont {Obry}, \citenamefont
  {Hillebrands},\ and\ \citenamefont
  {Schultheiss}}]{sebastian_micro-focused_2015}%
  \BibitemOpen
  \bibfield  {author} {\bibinfo {author} {\bibfnamefont {T.}~\bibnamefont
  {Sebastian}}, \bibinfo {author} {\bibfnamefont {K.}~\bibnamefont
  {Schultheiss}}, \bibinfo {author} {\bibfnamefont {B.}~\bibnamefont {Obry}},
  \bibinfo {author} {\bibfnamefont {B.}~\bibnamefont {Hillebrands}},\ and\
  \bibinfo {author} {\bibfnamefont {H.}~\bibnamefont {Schultheiss}},\
  }\bibfield  {journal} {\bibinfo  {journal} {Frontiers in Physics}\ }\textbf
  {\bibinfo {volume} {3}},\ \href {https://doi.org/10.3389/fphy.2015.00035}
  {10.3389/fphy.2015.00035} (\bibinfo {year} {2015})\BibitemShut {NoStop}%
\bibitem [{\citenamefont {Mock}\ \emph {et~al.}(1987)\citenamefont {Mock},
  \citenamefont {Hillebrands},\ and\ \citenamefont
  {Sandercock}}]{mock_construction_1987}%
  \BibitemOpen
  \bibfield  {author} {\bibinfo {author} {\bibfnamefont {R.}~\bibnamefont
  {Mock}}, \bibinfo {author} {\bibfnamefont {B.}~\bibnamefont {Hillebrands}},\
  and\ \bibinfo {author} {\bibfnamefont {R.}~\bibnamefont {Sandercock}},\
  }\href {https://doi.org/10.1088/0022-3735/20/6/017} {\bibfield  {journal}
  {\bibinfo  {journal} {Journal of Physics E: Scientific Instruments}\ }\textbf
  {\bibinfo {volume} {20}},\ \bibinfo {pages} {656} (\bibinfo {year}
  {1987})}\BibitemShut {NoStop}%
\bibitem [{\citenamefont {Tikui{\v{s}}is}\ \emph {et~al.}(2017)\citenamefont
  {Tikui{\v{s}}is}, \citenamefont {Beran}, \citenamefont {Cejpek},
  \citenamefont {Uhl{\'\i}{\v{r}}ov{\'a}}, \citenamefont {Hamrle},
  \citenamefont {Va{\v{n}}atka}, \citenamefont {Urb{\'a}nek},\ and\
  \citenamefont {Veis}}]{tikuivsis2017optical}%
  \BibitemOpen
  \bibfield  {author} {\bibinfo {author} {\bibfnamefont {K.~K.}\ \bibnamefont
  {Tikui{\v{s}}is}}, \bibinfo {author} {\bibfnamefont {L.}~\bibnamefont
  {Beran}}, \bibinfo {author} {\bibfnamefont {P.}~\bibnamefont {Cejpek}},
  \bibinfo {author} {\bibfnamefont {K.}~\bibnamefont
  {Uhl{\'\i}{\v{r}}ov{\'a}}}, \bibinfo {author} {\bibfnamefont
  {J.}~\bibnamefont {Hamrle}}, \bibinfo {author} {\bibfnamefont
  {M.}~\bibnamefont {Va{\v{n}}atka}}, \bibinfo {author} {\bibfnamefont
  {M.}~\bibnamefont {Urb{\'a}nek}},\ and\ \bibinfo {author} {\bibfnamefont
  {M.}~\bibnamefont {Veis}},\ }\href
  {https://doi.org/10.1016/j.matdes.2016.10.036} {\bibfield  {journal}
  {\bibinfo  {journal} {Materials \& Design}\ }\textbf {\bibinfo {volume}
  {114}},\ \bibinfo {pages} {31–39} (\bibinfo {year} {2017})}\BibitemShut
  {NoStop}%
\bibitem [{\citenamefont {Dieterle}\ \emph {et~al.}(2019)\citenamefont
  {Dieterle}, \citenamefont {F\"orster}, \citenamefont {Stoll}, \citenamefont
  {Semisalova}, \citenamefont {Finizio}, \citenamefont {Gangwar}, \citenamefont
  {Weigand}, \citenamefont {Noske}, \citenamefont {F\"ahnle}, \citenamefont
  {Bykova}, \citenamefont {Gr\"afe}, \citenamefont {Bozhko}, \citenamefont
  {Musiienko-Shmarova}, \citenamefont {Tiberkevich}, \citenamefont {Slavin},
  \citenamefont {Back}, \citenamefont {Raabe}, \citenamefont {Sch\"utz},\ and\
  \citenamefont {Wintz}}]{wintzHeterosymmetric}%
  \BibitemOpen
  \bibfield  {author} {\bibinfo {author} {\bibfnamefont {G.}~\bibnamefont
  {Dieterle}}, \bibinfo {author} {\bibfnamefont {J.}~\bibnamefont {F\"orster}},
  \bibinfo {author} {\bibfnamefont {H.}~\bibnamefont {Stoll}}, \bibinfo
  {author} {\bibfnamefont {A.~S.}\ \bibnamefont {Semisalova}}, \bibinfo
  {author} {\bibfnamefont {S.}~\bibnamefont {Finizio}}, \bibinfo {author}
  {\bibfnamefont {A.}~\bibnamefont {Gangwar}}, \bibinfo {author} {\bibfnamefont
  {M.}~\bibnamefont {Weigand}}, \bibinfo {author} {\bibfnamefont
  {M.}~\bibnamefont {Noske}}, \bibinfo {author} {\bibfnamefont
  {M.}~\bibnamefont {F\"ahnle}}, \bibinfo {author} {\bibfnamefont
  {I.}~\bibnamefont {Bykova}}, \bibinfo {author} {\bibfnamefont
  {J.}~\bibnamefont {Gr\"afe}}, \bibinfo {author} {\bibfnamefont {D.~A.}\
  \bibnamefont {Bozhko}}, \bibinfo {author} {\bibfnamefont {H.~Y.}\
  \bibnamefont {Musiienko-Shmarova}}, \bibinfo {author} {\bibfnamefont
  {V.}~\bibnamefont {Tiberkevich}}, \bibinfo {author} {\bibfnamefont {A.~N.}\
  \bibnamefont {Slavin}}, \bibinfo {author} {\bibfnamefont {C.~H.}\
  \bibnamefont {Back}}, \bibinfo {author} {\bibfnamefont {J.}~\bibnamefont
  {Raabe}}, \bibinfo {author} {\bibfnamefont {G.}~\bibnamefont {Sch\"utz}},\
  and\ \bibinfo {author} {\bibfnamefont {S.}~\bibnamefont {Wintz}},\ }\href
  {https://doi.org/10.1103/PhysRevLett.122.117202} {\bibfield  {journal}
  {\bibinfo  {journal} {Phys. Rev. Lett.}\ }\textbf {\bibinfo {volume} {122}},\
  \bibinfo {pages} {117202} (\bibinfo {year} {2019})}\BibitemShut {NoStop}%
\bibitem [{\citenamefont {Trevillian}\ and\ \citenamefont
  {Tyberkevych}(2024)}]{trevillian2024formation}%
  \BibitemOpen
  \bibfield  {author} {\bibinfo {author} {\bibfnamefont {C.}~\bibnamefont
  {Trevillian}}\ and\ \bibinfo {author} {\bibfnamefont {V.}~\bibnamefont
  {Tyberkevych}},\ }\bibfield  {journal} {\bibinfo  {journal} {npj
  Spintronics}\ }\textbf {\bibinfo {volume} {2}},\ \href
  {https://doi.org/10.1038/s44306-024-00026-3} {10.1038/s44306-024-00026-3}
  (\bibinfo {year} {2024})\BibitemShut {NoStop}%
\bibitem [{\citenamefont {Heins}\ \emph {et~al.}(2024)\citenamefont {Heins},
  \citenamefont {Iurchuk}, \citenamefont {Gladii}, \citenamefont {Körber},
  \citenamefont {Kakay}, \citenamefont {Faßbender}, \citenamefont
  {Schultheiß},\ and\ \citenamefont
  {Schultheiß}}]{heins_christopher_2024_3336}%
  \BibitemOpen
  \bibfield  {author} {\bibinfo {author} {\bibfnamefont {C.}~\bibnamefont
  {Heins}}, \bibinfo {author} {\bibfnamefont {V.}~\bibnamefont {Iurchuk}},
  \bibinfo {author} {\bibfnamefont {O.}~\bibnamefont {Gladii}}, \bibinfo
  {author} {\bibfnamefont {L.}~\bibnamefont {Körber}}, \bibinfo {author}
  {\bibfnamefont {A.}~\bibnamefont {Kakay}}, \bibinfo {author} {\bibfnamefont
  {J.}~\bibnamefont {Faßbender}}, \bibinfo {author} {\bibfnamefont
  {K.}~\bibnamefont {Schultheiß}},\ and\ \bibinfo {author} {\bibfnamefont
  {H.}~\bibnamefont {Schultheiß}},\ }\href
  {https://doi.org/10.14278/rodare.3336} {\bibinfo {title} {{Data publication:
  Nonreciprocal spin-wave dispersion in magnetic bilayers}}} (\bibinfo {year}
  {2024})\BibitemShut {NoStop}%
\end{thebibliography}%
\bibliographystyle{apsrev4-2}
\end{document}